\documentclass[english,a4paper]{article}

\usepackage{amsmath}
\usepackage{amssymb}
\usepackage[margin=2.5cm]{geometry}
\usepackage{authblk}

\usepackage[english]{babel}
\usepackage{graphicx}

\usepackage{url}
\usepackage{xcolor}

\usepackage{listings}
\renewcommand{\lstlistingname}{List.}
\renewcommand{\lstlistingname}{Listing}
\usepackage[capitalise,nameinlink]{cleveref}
  \crefname{section}{Sect.}{Sect.}
  \Crefname{section}{Section}{Sections}
  \crefname{listing}{\lstlistingname}{\lstlistingname}
  \Crefname{listing}{Listing}{Listings}

\usepackage{graphbox} 

\graphicspath{{figures/}}
\usepackage[multidot]{grffile}
\usepackage[caption=false]{subfig}

\DeclareCaptionFormat{mylst}{\hrule#1#2#3}
\newcommand{\dist}[1]{\ensuremath{\text{dist}\mkern-3mu\left({#1}\right)}}
\newcommand{\norm}[1]{\ensuremath{\left\|{#1}\right\|}}
\newcommand{\setsize}[1]{\ensuremath{\left|{#1}\right|}}
\newcommand{\len}[1]{\ensuremath{\text{len}\mkern-3mu\left({#1}\right)}}
\newcommand{\intlen}[2]{\ensuremath{\text{len}_{#1}\mkern-3mu\left({#2}\right)}}
\renewcommand{\Re}{\mathbb{R}}
\newcommand{\gm}{\gamma}
\newcommand{\EE}{{\mathbb E}}
\newcommand{\R}{\Re}

\newcommand{\dd}{\mathrm{d}}
\newcommand{\ld}{\lambda}
\newcommand{\af}{\alpha}
\newcommand{\jon}[1]{[Jon says: \textbf{#1}]}
\renewcommand{\jon}[1]{}

\newcommand{\rwrev}[1]{{\color{magenta}{#1}}}
\renewcommand{\rwrev}[1]{#1}

\begin{document}
\title{Generalizations of Ripley’s $K$-function with Application to Space Curves}
\author[*]{Jon Sporring}
\author[**]{Rasmus Waagepetersen}
\author[*]{Stefan Sommer}
 \affil[*]{Department of Computer Science, University of Copenhagen, Denmark\newline\{sporring,sommer\}@di.ku.dk}
\affil[**]{Department of Mathematical Sciences, Aalborg University, Denmark\newline rw@math.aau.dk}

\maketitle

\begin{abstract}
  The intensity function and Ripley's $K$-function have been used extensively in the literature to describe the first and second moment structure of spatial point sets. This has many applications including describing the statistical structure of synaptic vesicles. Some attempts have been made to extend Ripley's $K$-function to curve pieces. Such an extension can be used to describe the statistical structure of muscle fibers and brain fiber tracks. In this paper, we take a computational perspective and construct new and very general variants of Ripley's $K$-function for curves pieces, surface patches etc. We discuss the method from \cite{chiu:stoyan:kendall:mecke:13} and compare it with our generalizations theoretically, and we give examples demonstrating the difference in their ability to separate sets of curve pieces.
\end{abstract}

\section{Introduction}
\label{sec:intro}
\jon{Remember to acknowledge CSGB and the Villum foundation.}  In this paper, we consider descriptive statistics of sets of curves and particularly whether or not there \rwrev{is a tendency} for the curves to cluster or repel each other. In the human body, such curve structures appear in multiple places. Examples include skeletal muscles and fiber tracts in the human brain. We here aim to introduce descriptive statistics for the medical imaging community and discuss estimation methods for geometric data using medical imaging analysis techniques such as morphology and currents.

Ripley's $K$-function \cite{ripley77} is a well-established tool for describing the second moment structure of point sets \cite{baddeley:rubak:turner:15}, and some attempt has been made to generalize this function to curve pieces as described in \cite{chiu:stoyan:kendall:mecke:13}.

For a homogeneous point point pattern $p_i \in \Re^d$, $i=1\ldots n$ in an observation window $W$, the sample-based estimate of Ripley's $K$-function \cite{ripley77} measures the deviation from \rwrev{complete spatial randomness. Complete spatial randomness informally means that points occur uniformly in space and independently of each other. More formally, the points are distributed according to a Poisson point process with constant intensity.} An estimator of Ripley's $K$-function is,
\begin{equation}
  \label{eq:ripleysK}
  K(r) = \frac{1}{n\lambda}\sum_{i\neq j}1(\dist{p_i,p_j}<r),
\end{equation}
where $n$ is the number of points, $\lambda = \frac{\setsize{W}}{n}$ is the \rwrev{sample intensity}, and $1$ is the indicator function. Deviations from homogeneity are identified by comparing $K(r)$ with the volume of a $d$-dimensional ball of radius $r$, i.e., if $K(r)$ is greater than or smaller than the volume of the said ball, then points tend to cluster or repel each other respectively at \rwrev{distances $r$}.

\begin{figure}%
  \centering
  \subfloat[Random]{\label{fig:RipleysKOnPoints:uniform}\includegraphics[width=0.45\linewidth]{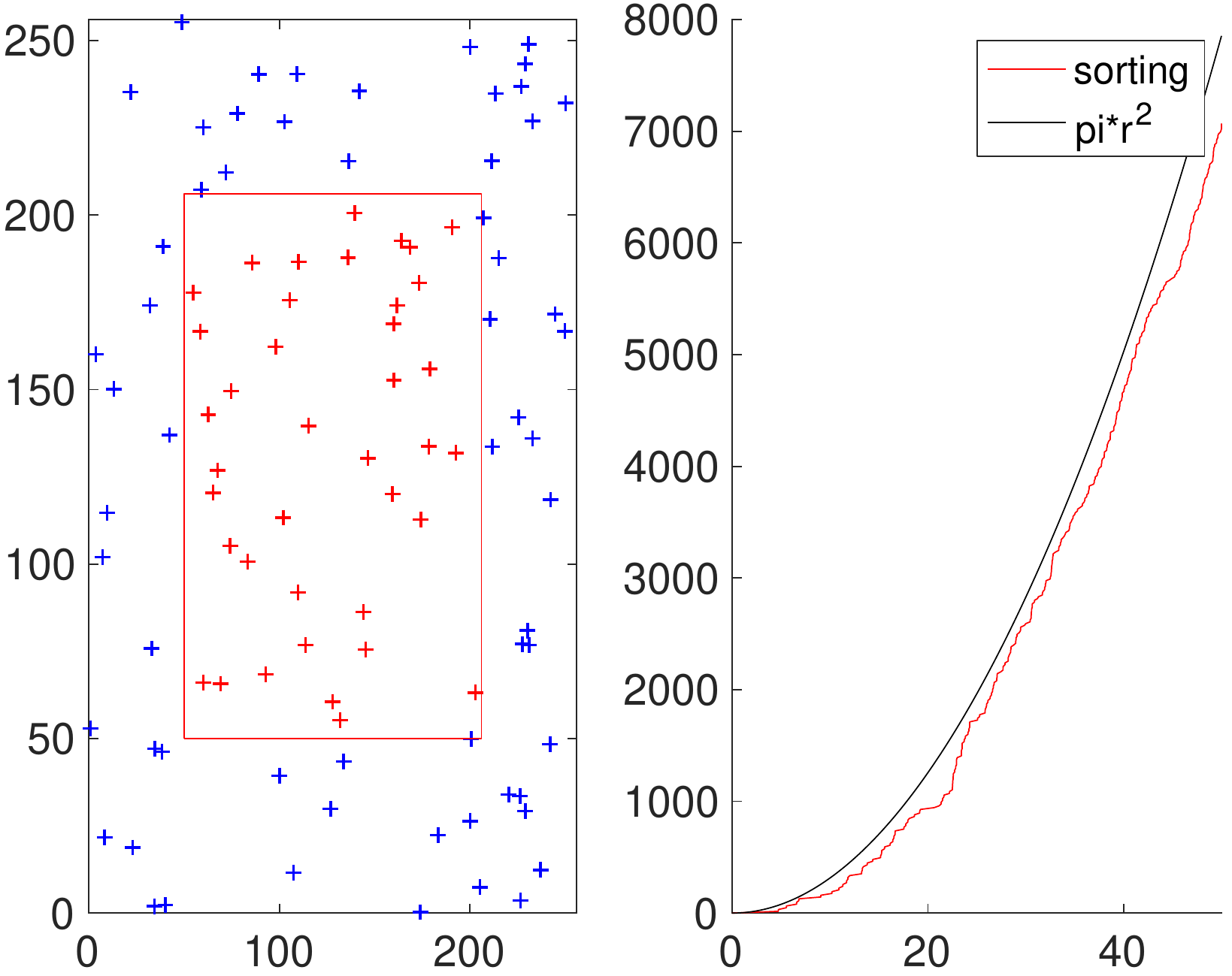}}\hspace{1mm}
  \subfloat[Local clustering]{\label{fig:RipleysKOnPoints:motherofgauss}\includegraphics[width=0.45\linewidth]{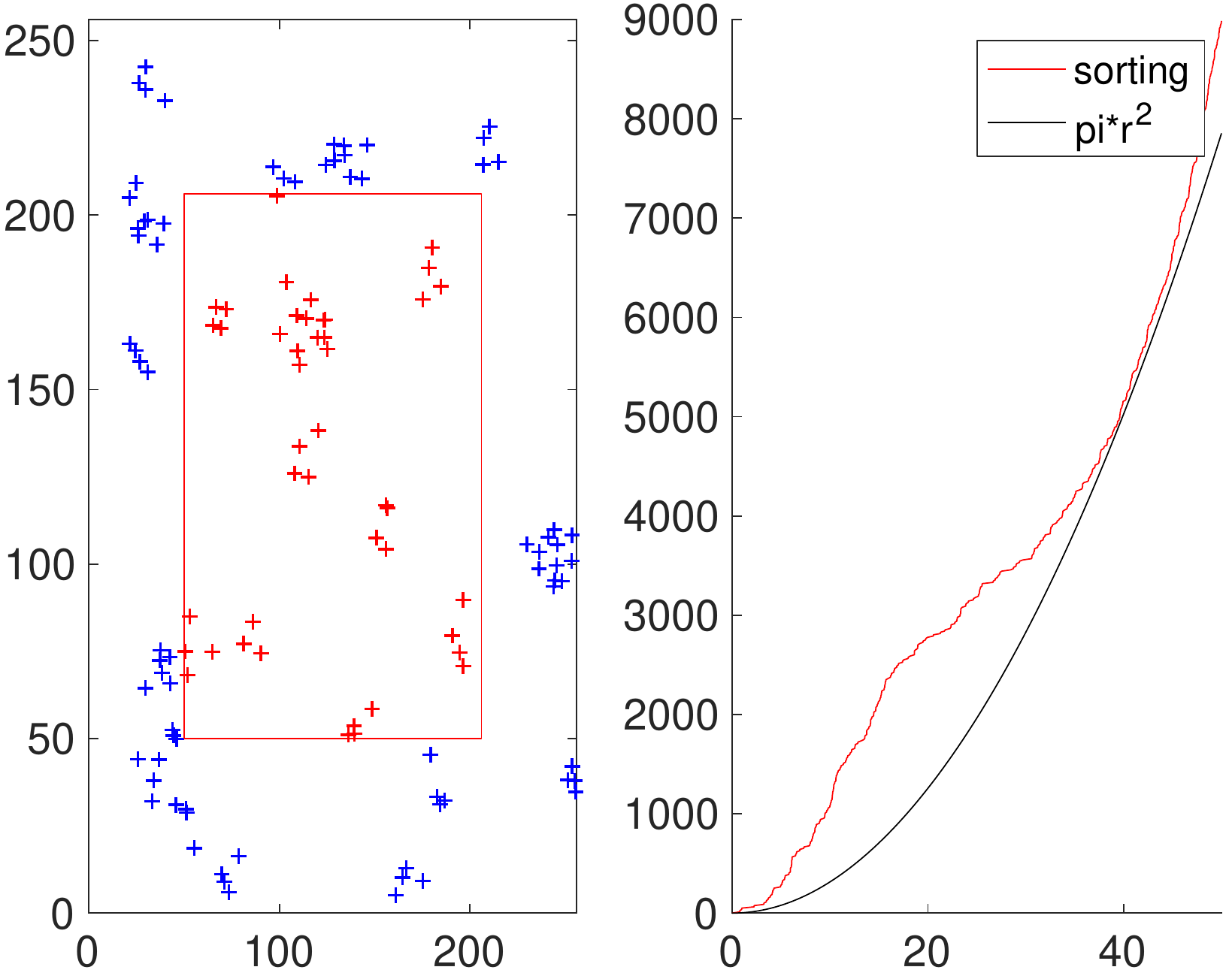}}\hspace{1mm}
  \subfloat[Noisy grid]{\label{fig:RipleysKOnPoints:noisygrid}\includegraphics[width=0.45\linewidth]{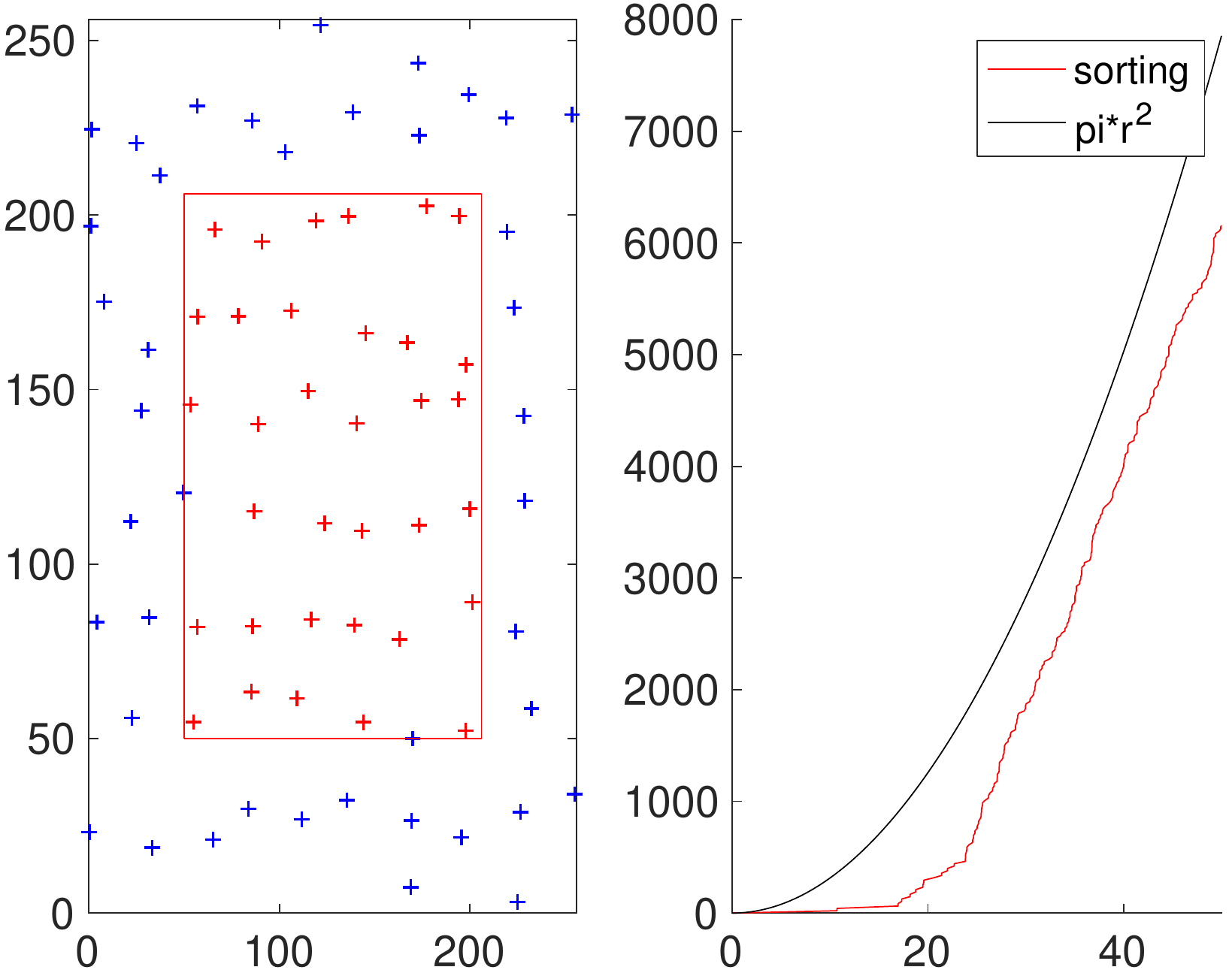}}\hspace{1mm}
  \subfloat[Regular grid]{\label{fig:RipleysKOnPoints:grid}\includegraphics[width=0.45\linewidth]{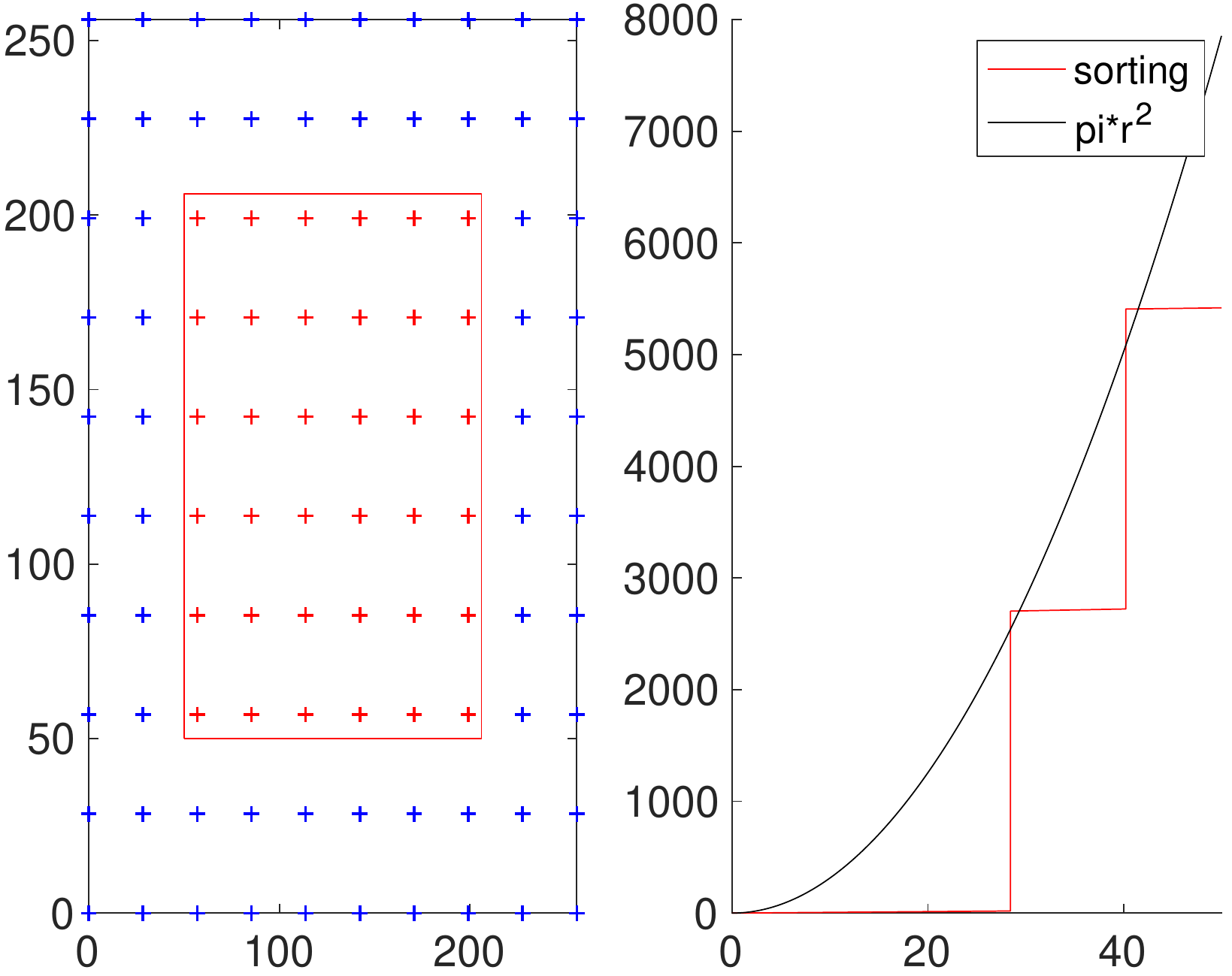}}
  \caption{Point sets and Ripley's $K$-function estimated by \Cref{lst:sorting} and compared with the theoretical function for \rwrev{complete spatial randomness}. The red square denotes the observation window. The figures show the estimation for random point sets that are (a) independently and uniformly distributed, (b) randomly generated mother processes with independent and normally distributed children, (c) regular grid points perturbed by a normal distribution centered in the gridpoint, and (d) regular grid points.}
  \label{fig:RipleysKOnPoints}
\end{figure}

For this article, we highlight the following algorithm for computing \eqref{eq:ripleysK}. Given a set of points $\{p_i\}$, we estimate Ripley's $K$-function by the following algorithm.
\begin{lstlisting}[mathescape,caption={An algorithm for estimating Ripley's $K$-function by sorting},label={lst:sorting},float]
// Calculate the distance matrix 
1. $d_{ij} \leftarrow \dist{p_i,p_j}$
// Sort the non-diagonal values of $d_{ij}$ in increasing order.
2. $r_k \leftarrow \text{sort}(d_{ij}), \text{ for } i\neq j \text{ and } k=1,2,\ldots (n^2 - n)$
// Generate the set of pairs ordered by $k$
3. $c \leftarrow \{c_k = (r_k,k)\}$
// Remove duplicates of $r_k$ keeping that with highest value of $k$
4. do 
5.   $\text{del} \leftarrow \{(r_k,k) | \exists (r_k', k'), \text{ where } r_k = r_k' \text{ and } k < k') \}$
6.   $c \leftarrow c \setminus \text{del}$
7. until $c = \{\}$
// Assign values to estimator (non-uniformly sampled).
8. $K(r_l) \leftarrow l$
\end{lstlisting}
Examples of using this algorithm on different point sets in $\Re^2$ are given in \Cref{fig:RipleysKOnPoints}.
The figure shows 4 different randomly generated point sets together with the estimated Ripley's $K$-function. The red squares denotes the observation window. In this article, we will not discuss boundary artifacts and for this reason, we include the blue points asymmetrically, such that when evaluating $\dist{p_i,p_j}$, we let $i$ iterate over all red points and $j$ over all red and blue points except where $i=j$. The black curve is the theoretical value for Ripley's $K$-function \rwrev{under complete spatial randomness}. In the figures, we see that for randomly and uniformly distributed points, as in \Cref{fig:RipleysKOnPoints:uniform}, the estimator tend to approximate the curve $\pi r^2$. For locally clustered points, as in \Cref{fig:RipleysKOnPoints:motherofgauss}, the estimator lies above $\pi r^2$ for a range of distances approximately corresponding to the width of the local cluster. For noisy grid points \Cref{fig:RipleysKOnPoints:noisygrid} and grid points \Cref{fig:RipleysKOnPoints:grid}, the estimator lies below $\pi r^2$.

Ripley's $K$-function may be extended to more general random sets of geometrical objects than point sets. A $K$-function for so-called fiber-processes is reviewed in \cite{chiu:stoyan:kendall:mecke:13}.

Related work is \cite{Durrleman.Pennec.Trouve.Ayache09}, where currents are used to define moments of sets of curves and other geometrical primitives. This allows the authors to visualize and build simplified models using the principal modes of the covariance structure.

In this article, we propose two extensions of Ripley's $K$-function for manifold pieces of any dimension, which includes the original definition for point sets, and compare \cite{chiu:stoyan:kendall:mecke:13} with our alternatives theoretically and on simple sets.

\section{$K$-function for fibers: $K_f$}
\label{sec:chiu}
In \rwrev{\cite{chiu:stoyan:kendall:mecke:13}, is discussed a number of spatial descriptive statistics including a $K$-function for curve segments, coined fibers.}  A fiber process $\Phi \subseteq \R^d$ is a random union of fibers (or curve segments) $\gm$ where each $\gm$ satisfies that the length of $\gm \cap B$ is well-defined and finite for all bounded $B \subset \R^d$. Moreover, for any bounded $B \subseteq \R^d$, the random total length $L(B)$ of fibers in $\Phi \cap B$ is finite. The current approach, e.g., reviewed in \cite[Chapter 8]{chiu:stoyan:kendall:mecke:13} to statistically characterizing fiber processes is based on moment properties of the random variables $L(B)$ for bounded subsets $B\subset \R^d$.

Assuming that $\Phi$ is stationary (i.e.\ its distribution is invariant to translations), the intensity $\rho$ of $\Phi$ is defined by 
\[ \rho = \frac{\EE L(B)}{|B|} \]
where $B$ is any subset of positive $d$-dimensional volume $|B|$. In other words, $\rho$ is the expected length of fibers in a set of unit volume.

The second moment measure is defined via second order moments of the length variables:
\[ \mu^{(2)}(A \times B)= \EE[ L(A) L(B) ] \]
for bounded $A, B \subset \R^d$. By standard measure-theoretical results, $\mu^{(2)}$ can be decomposed as
\[ \mu^{(2)}(A \times B) = \rho^2 \int_A {\cal K}(B-x) \dd x \]
where ${\cal K}(\cdot)$ defined on subsets of $\R^d$ is called the reduced moment measure. Here $B-x$ is $B$ translated by the vector $-x$ for $B \subset \R^d$ and $x \in \R^d$.

Letting $b(0,r)$ denote the $d$-dimensional ball of radius $r$ and
centered in the origo, we obtain the $K$-function \rwrev{for fibers}
\[ K_f(r)={\cal K}(b(0,r)). \]
For each $r \ge 0$, $K_f(r)$ can be interpreted a the expected total fiber length in $b(u,r)$ conditional on that $u \in \Phi$ for $u \in \R^d$. Thus, given that a fiber intersects $u \in \R^d$, the $K_f$-function describes the tendency of further fibers in $\Phi$ to aggregate in or avoid the ball $b(u,r)$ centered around $u$.

\subsection{Relation to Cox process}
\label{sec:coxProcess}
Suppose we generate a point process $Y$ by placing points on $\Phi$ according to a Poisson process. More specifically, for each fiber $\gm$ in $\Phi$, a Poisson process of intensity $\lambda$ is generated on $\gm$. Then the intensity of $Y$ becomes $\lambda \rho$.  The second order factorial moment measure of $Y$ is closely related to the second moment measure of $\Phi$:
\begin{align*}
  \af^{(2)}_Y(A \times B) & =\EE \sum_{u,v \in Y}^{\neq}
                            1[u \in A, v \in B]  = \ld^2
                            \EE \int_\Phi \int_\Phi 1[u
                            \in A, v \in B] \dd u \dd v
  \\ & =  \ld^2 \EE[ \Phi(A) \Phi(B)]  = \ld^2 \mu^{(2)} (A \times B)
\end{align*}
for bounded $A, B\subset \R^d$.  Here and in the following $\int_\Phi$ should be interpreted as a curve integral along the fibers in $\Phi$.

The usual point process $K$-function for $Y$ is 
\[ K_Y(r)= {\cal K}_Y(b(0,r))\]
where ${\cal K}_Y(b(0,r))$ is defined by the equation
\[ \af^{(2)}_Y(A \times B)= \ld^2 \int_A {\cal K}_Y(B-x) \dd x .\]
We thus obtain that the $K$-functions of $\Phi$ and $Y$ coincide,
\[ K_f(r)=K_Y(r) \]
and 
 \begin{align*} 
 \rho K_f(r) & = \frac{1}{\rho |A|} \EE \left[ \int_\Phi \int_\Phi 1[ u \in A, \|v-u\| \le r] \dd v \dd u \right ]\\
 & = \EE \left[ \frac{1}{\rho |A|}  \int_{\Phi \cap A} L(b(u,r)) \dd u \right ].
 \end{align*}
 It follows that $\rho K_f(r)$ can further be viewed as the expectation of a spatial average of lengths $L(b(u,r))$ for $u \in \Phi \cap A$.

\subsection{Estimation of $K_f$}
\begin{figure}%
  \centering
  \subfloat[Point distance]{{\label{fig:geometricComparison:Points}\includegraphics[align=t,width=0.28\linewidth,viewport= 500 35 900 430,clip]{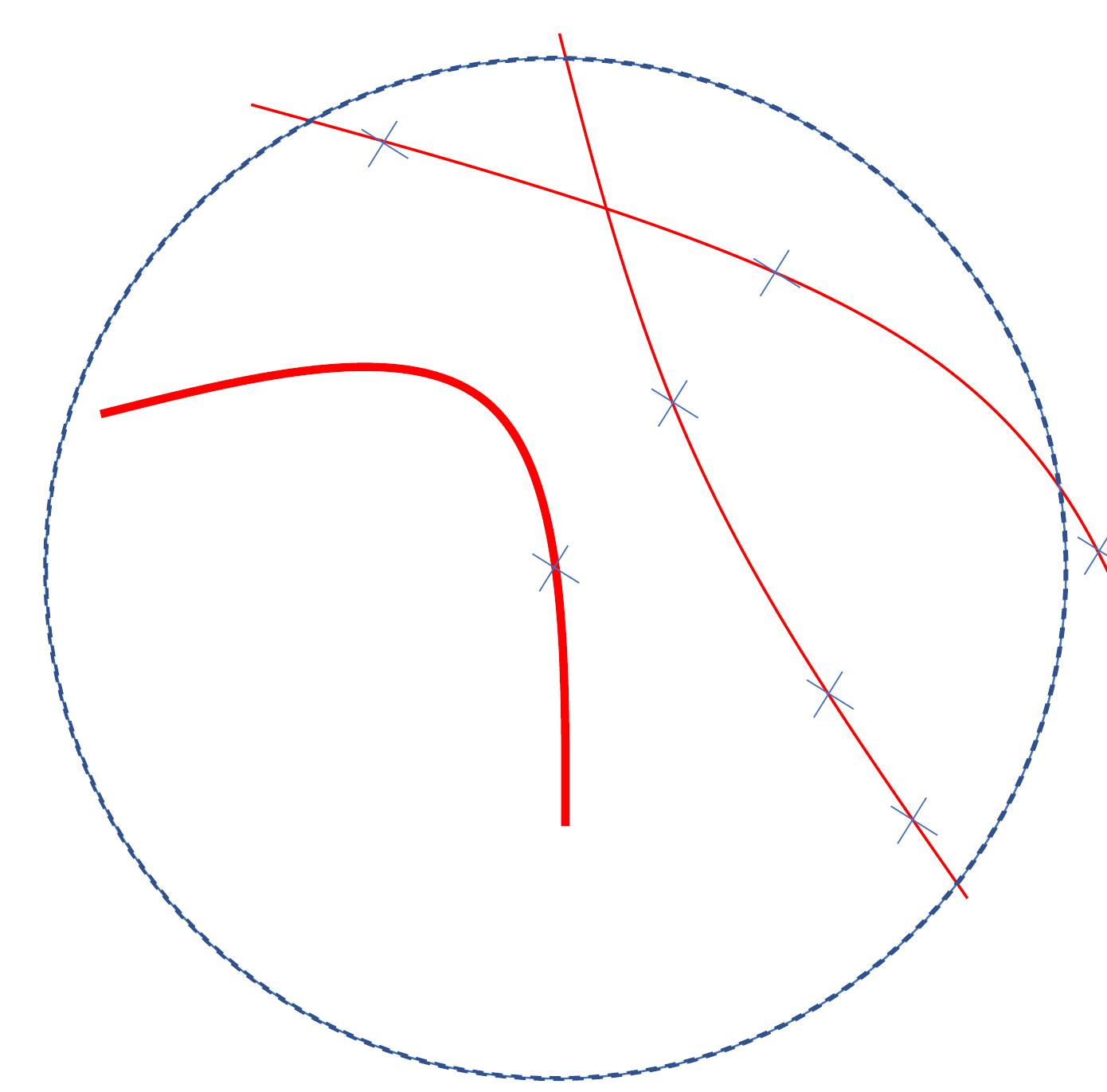}}}\hspace{1mm}
  \subfloat[Dilation distance]{\label{fig:geometricComparison:Dilation}\raisebox{-2.5mm}{\includegraphics[align=t,width=0.33\linewidth,viewport= 480 150 940 400,clip]{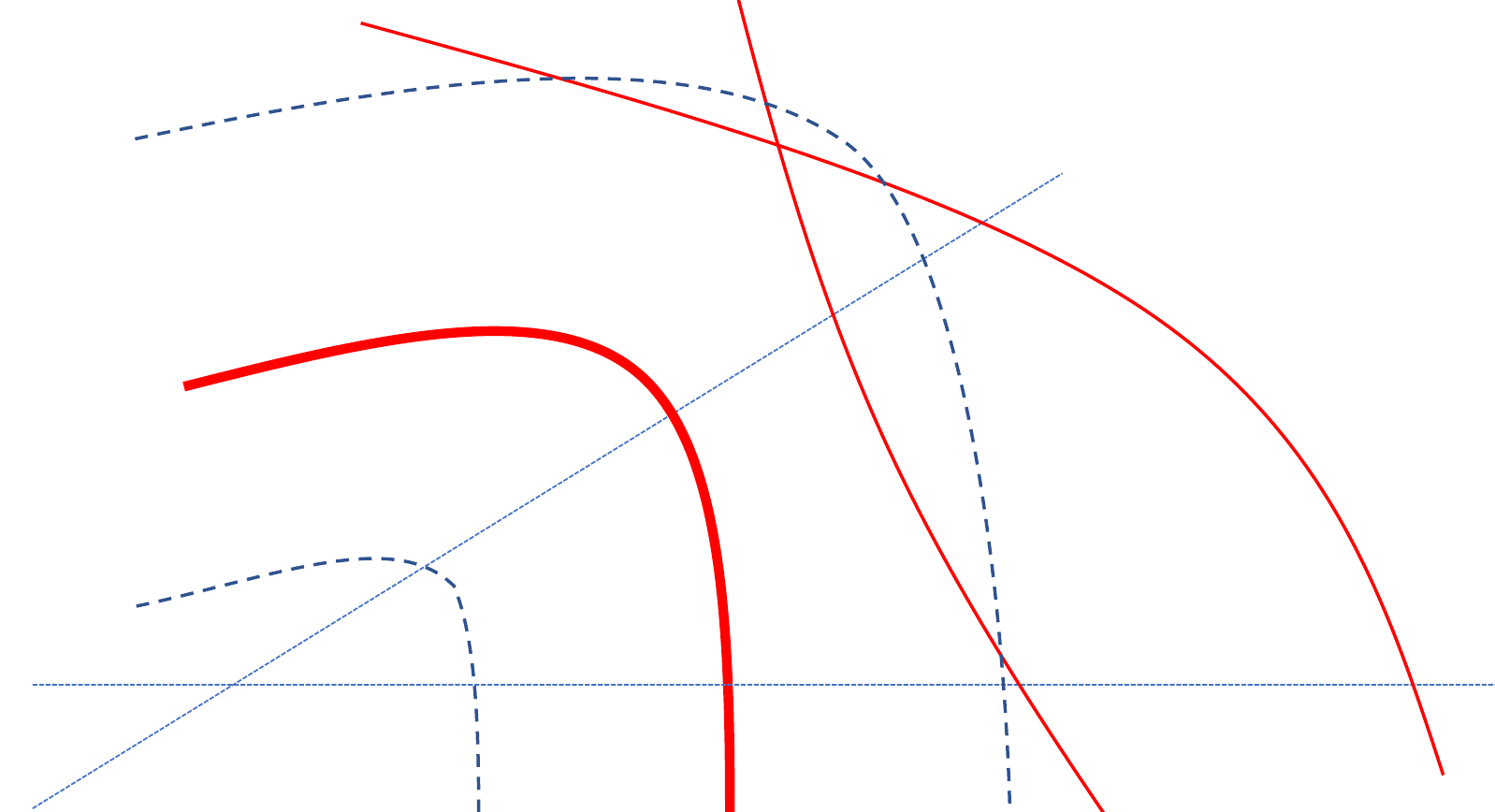}}}\hspace{1mm}
  \subfloat[Current distance]{{\label{fig:geometricComparison:Currents}\includegraphics[align=t,width=0.28\linewidth,viewport= 500 35 900 430,clip]{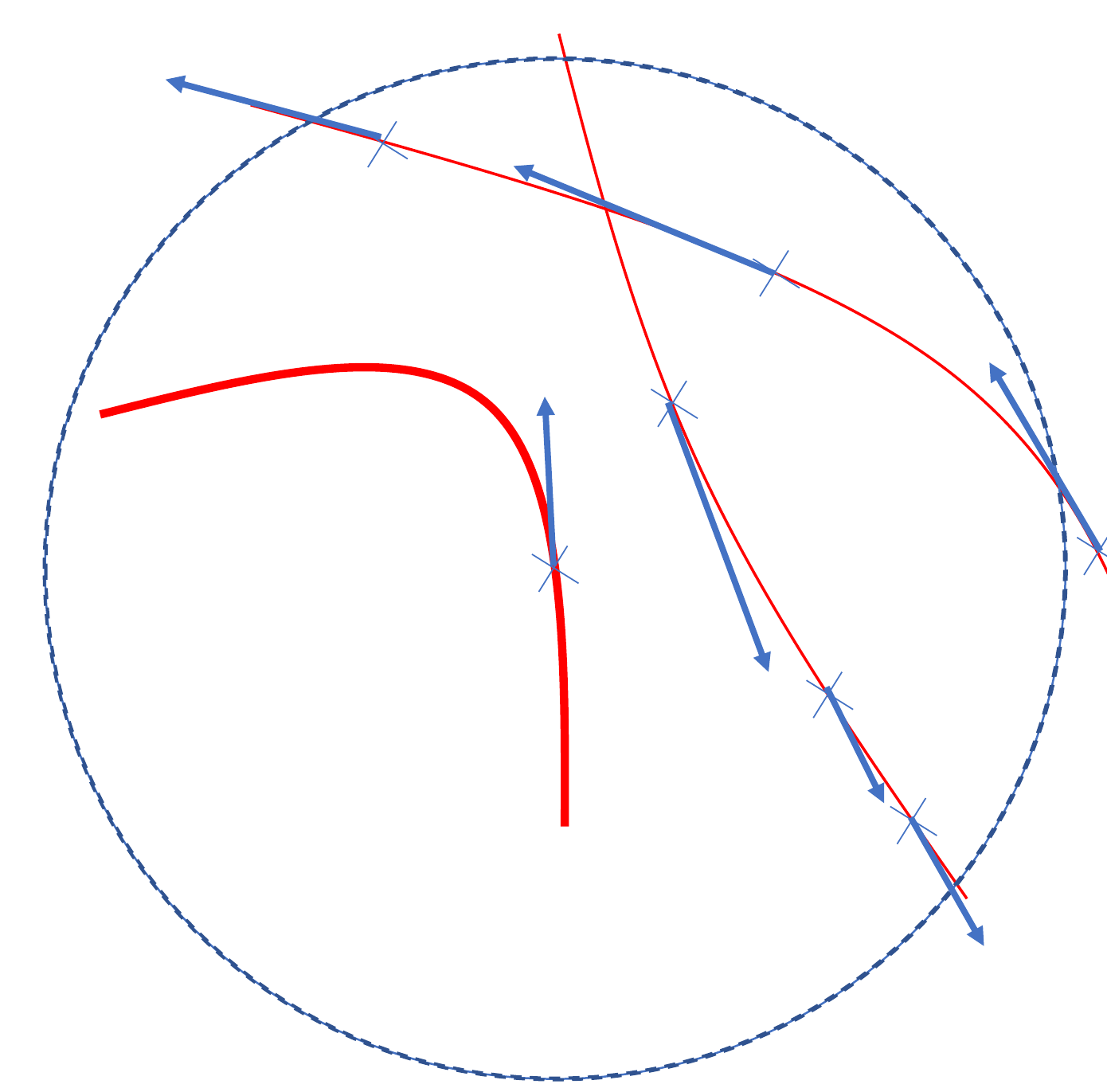}}}
  \caption{Geometric comparison of three different $K$-estimators. (a) $K_f$ discussed in \Cref{sec:coxProcess}, (b) $K_m$ discussed in \Cref{sec:morphology}, and (c) $K_c$ discussed in \Cref{sec:currents}.}
  \label{fig:geometricComparison}
\end{figure}

Suppose $\Phi$ is observed within a bounded observation window $W \subset \R^d$ and let $W\ominus r = \{ x \in W| b(x,r) \subset W\}$ denote the erosion of $W$ by the distance $r$. Then
\[  \hat K_f(r) = \frac{1}{\rho^2 |W \ominus r|} \int_{\Phi \cap W\ominus r} L(b(u,r)) \dd u  \]
is an unbiased estimate of $K_f$. Alternatively, one can generate a
point process $Y$ on $\Phi$ as described in the previous section and
estimate $K_f(r)=K_Y(r)$ by the usual estimator of point process
$K$-functions \rwrev{as described in Section~\ref{sec:intro}}. This essentially corresponds to evaluating the integrals/lengths in $\hat K_f(r)$ by Monte Carlo integration. \Cref{fig:geometricComparison:Points} shows an illustration of this approximation.
The figure shows a step in the process of estimating $K_f$: a ball has been placed around a point on a curve (thick line), and the contribution to the estimate for that radius and that point is the number of other points on the other lines.

\subsection{$K_f$ in terms of curve segments}
Let $\Phi=\cup_{i=1}^\infty \gamma_i$ where the $\gamma_i$ represent the individual fibers/curve segments and let for a bounded $A$,
 \[ d_A(\gm_i,\gm_j) = \int_{\gm_i \cap A} \int_{\gm_j} 1[ \|v-u\| \le r] \dd v
   \dd u. \]
Thus, a large value of $d_A(\gm_i,\gm_j)$ indicates that $\gm_i \cap A$ and $\gm_j$ are `close' so that $d_A$ is a kind of measure of association between $\gm_i$ and $\gm_j$. Then
\begin{align*}
  \rho K_f(r)  & = \frac{1}{\rho |A|} \EE \left[ \sum_{\gm_i:
                 \gm_i \cap A \neq \emptyset} \sum_{\gm_j}  \int_{\gm_i \cap A}
                 \int_{\gm_j} 1[ \|v-u\| \le r] \dd v \dd u \right ]\\
               & = \frac{1}{\rho |A|} \EE \left[ \sum_{\gm_i:
                 \gm_i \cap A \neq \emptyset} \sum_{\gm_j} d_A(\gm_i,\gm_j) \right ]
\end{align*}
Thus, $K_f$ can be viewed as a kind of average of $d_A(\gm_i,\gm_j)$ for pairs of distinct fibers $\gm_i$ and $\gm_j$ in $\Phi$. In this paper, we exploit this point of view by investigating alternative measures of association between fibers.

\subsection{Some Experiments with Random Curves and $K_f$}
\label{sec:ExperimentsChiu}
We will in the following motivate alternative definitions of $K_f$ by presenting a number of comparable cases in two dimensions.

\begin{figure}%
  \centering
  \includegraphics[width=0.35\linewidth]{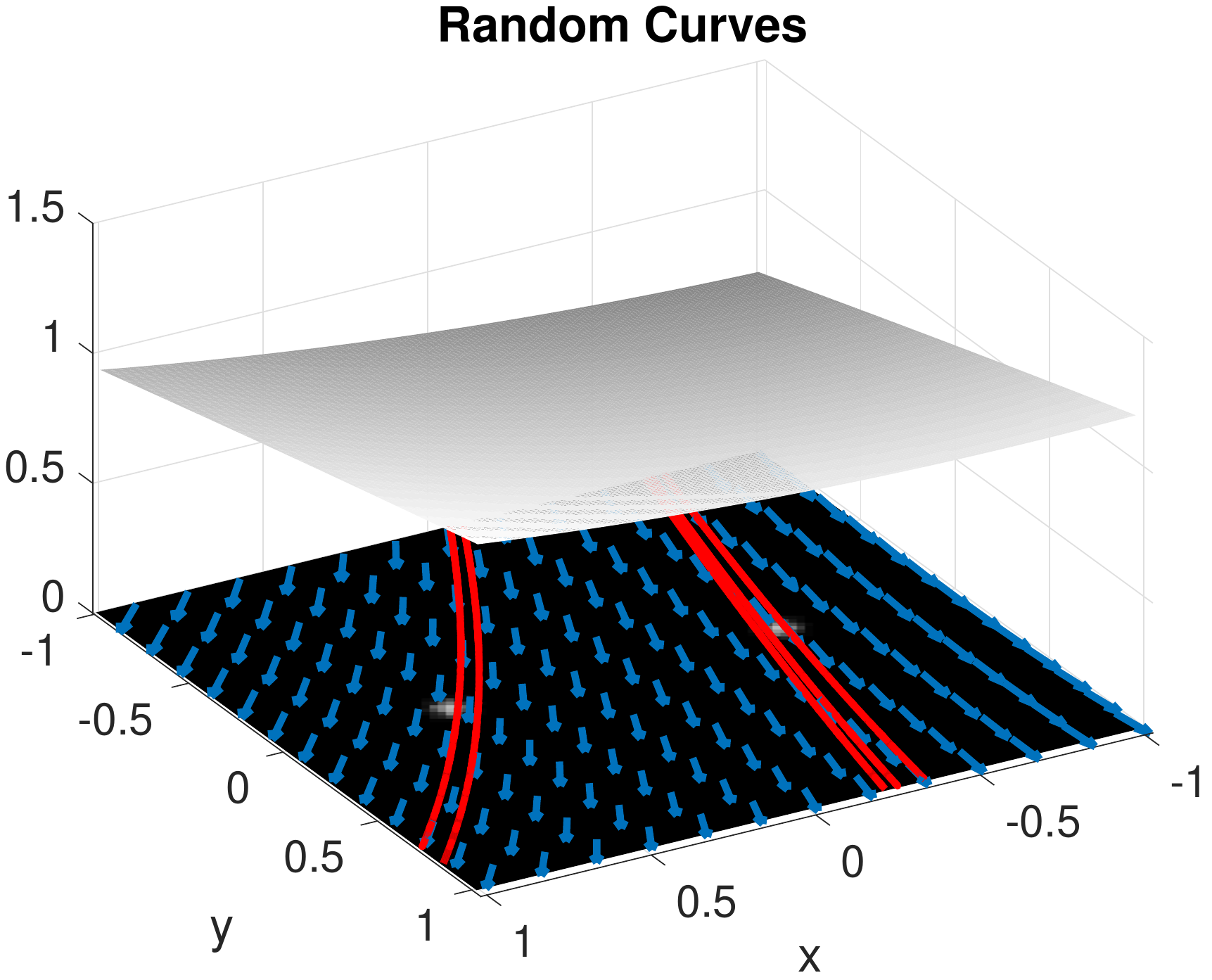}
  \caption{An example of a randomly generated set of curves: A random surface is shown in gray hovering above the lines, its gradient field is shown as blue arrows, a distribution of starting points is shown as a gray image, and red lines are flow lines each passing through random starting point.}
  \label{fig:generatingRandomCurves}
\end{figure}

\begin{figure}%
  \centering
  \subfloat[Random]{\includegraphics[width=0.23\linewidth]{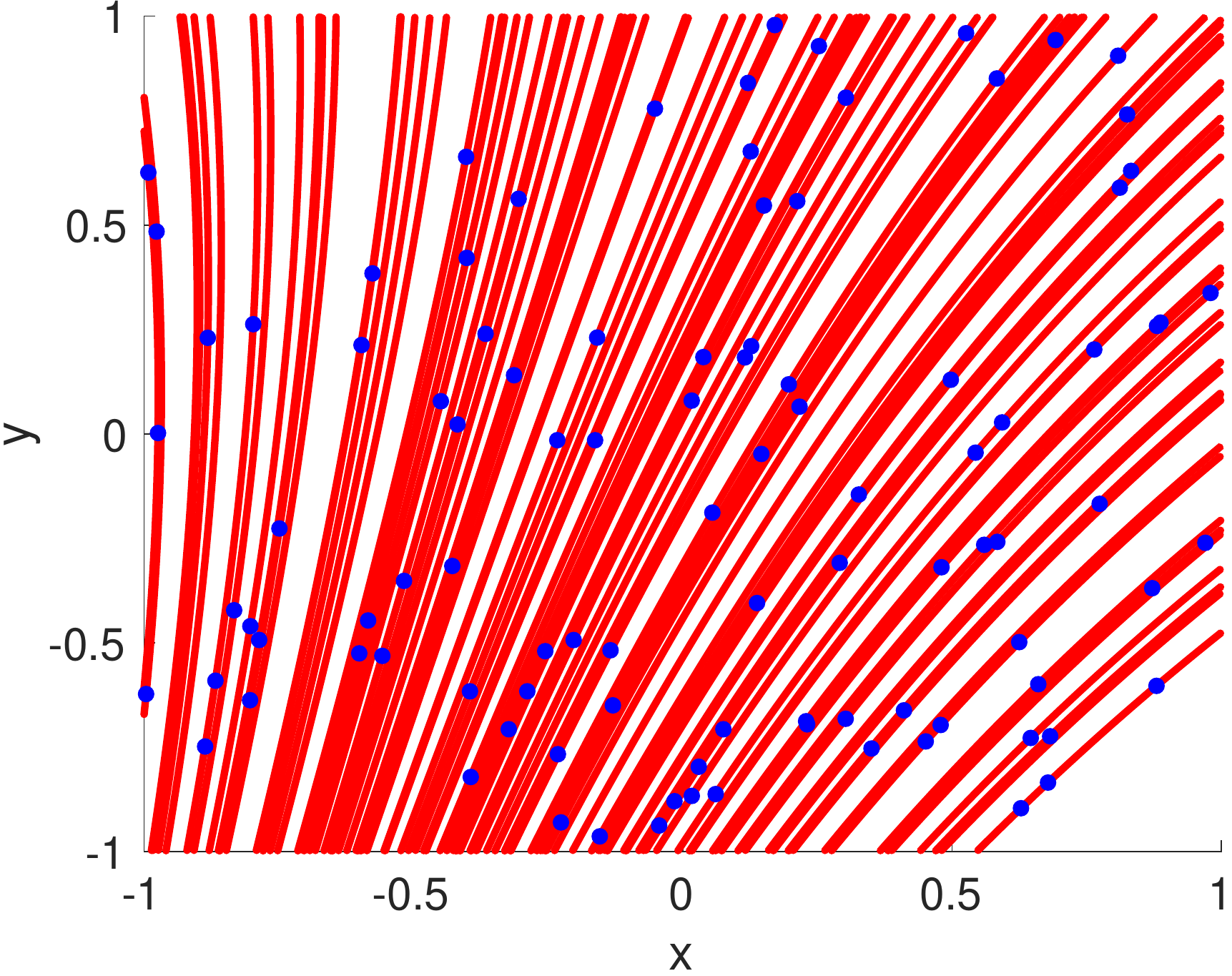}}\hspace{1mm}
  \subfloat[7 cluster]{\includegraphics[width=0.23\linewidth]{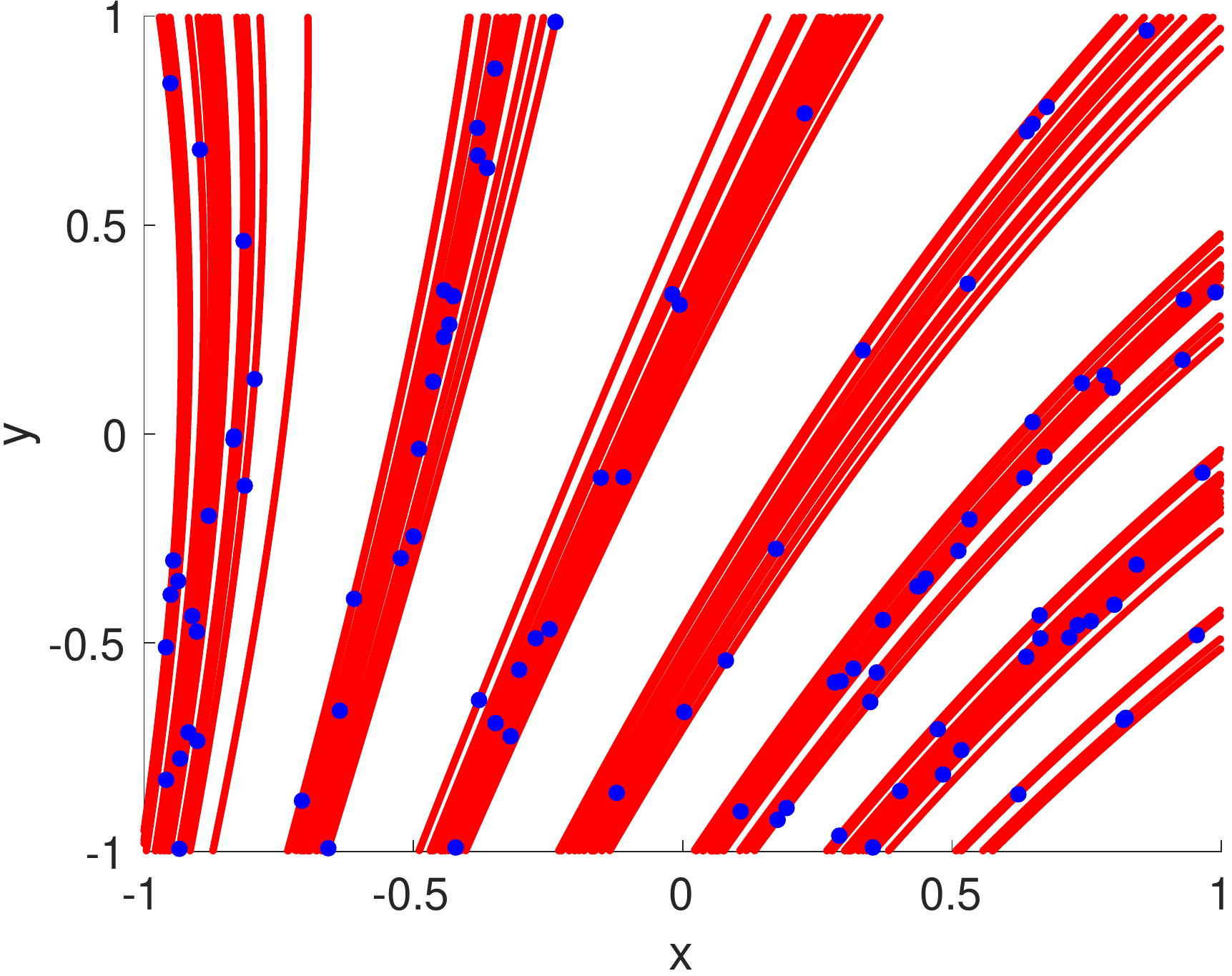}}\hspace{1mm}
  \subfloat[2 clusters]{\includegraphics[width=0.23\linewidth]{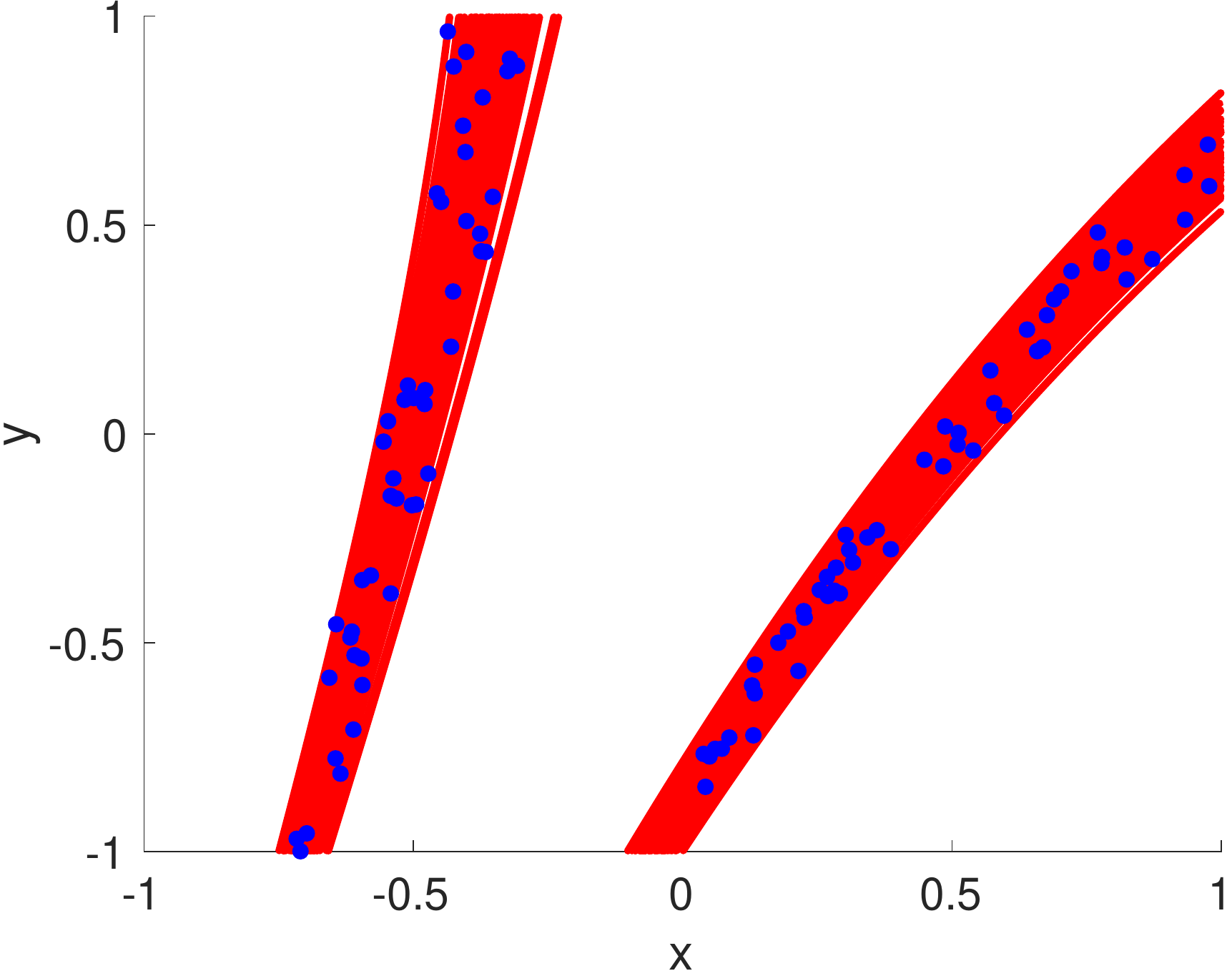}}\hspace{1mm}
  \subfloat[1 clusters]{\includegraphics[width=0.23\linewidth]{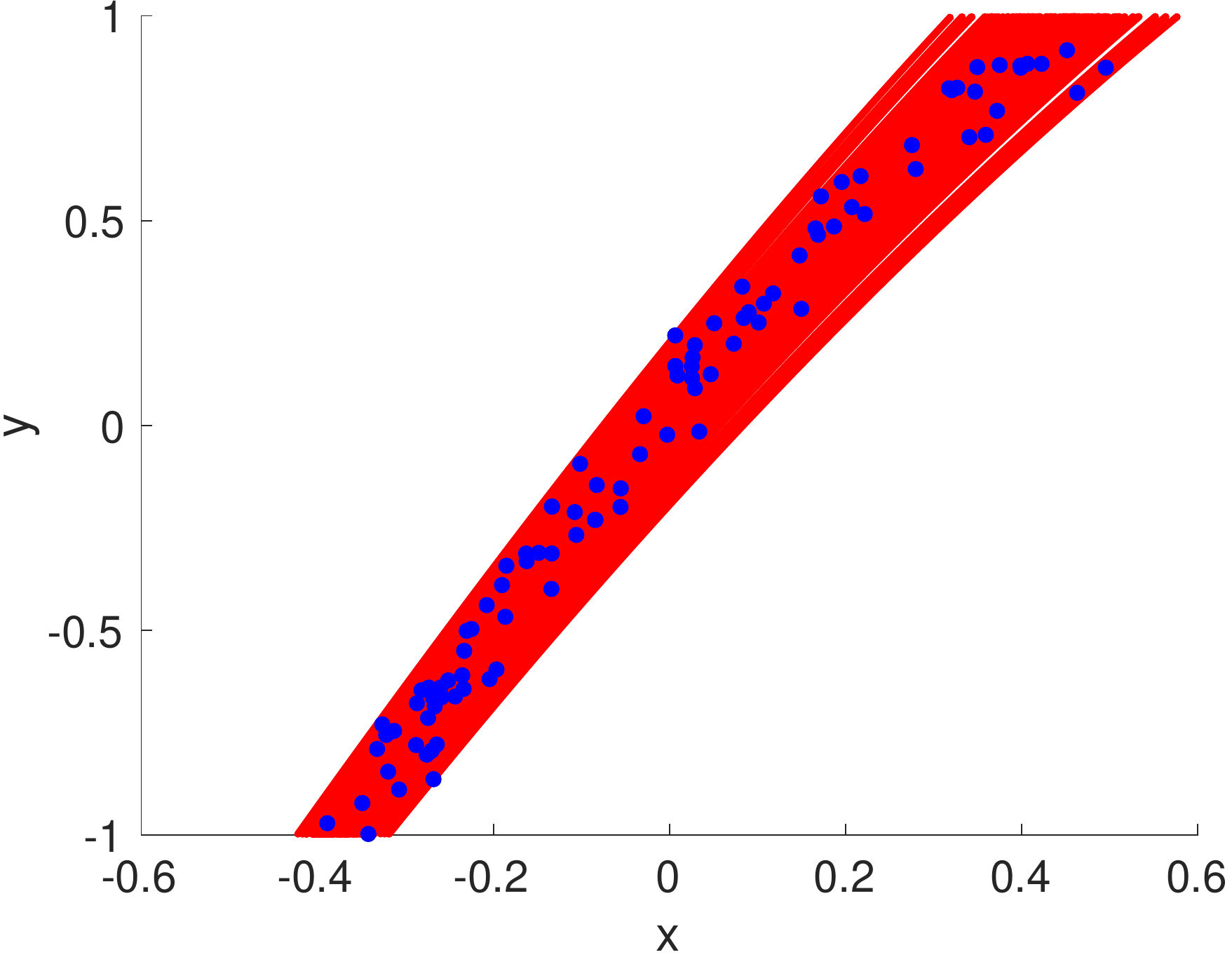}}
  \caption{Random curves and random points on curves. All curves are generated from the same vector field shown in \Cref{fig:generatingRandomCurves}.}
  \label{fig:randomPoints}
\end{figure}

\begin{figure}%
  \centering
  \subfloat[Wide]{\label{fig:randomPointsKWide}\includegraphics[width=0.23\linewidth]{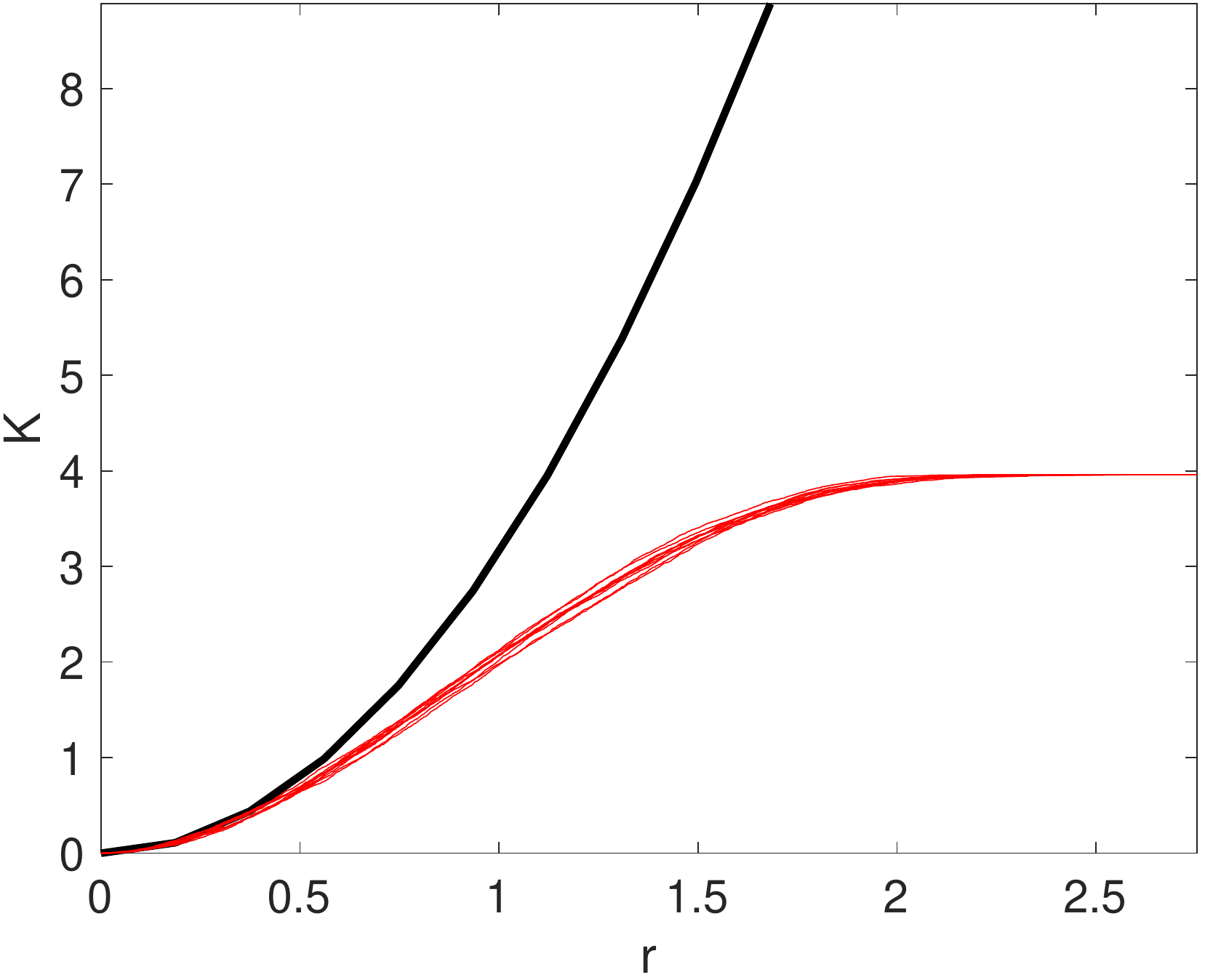}}\hspace{1mm}
  \subfloat[7 narrow clusters]{\label{fig:randomPointsK7Narrow}\includegraphics[width=0.23\linewidth]{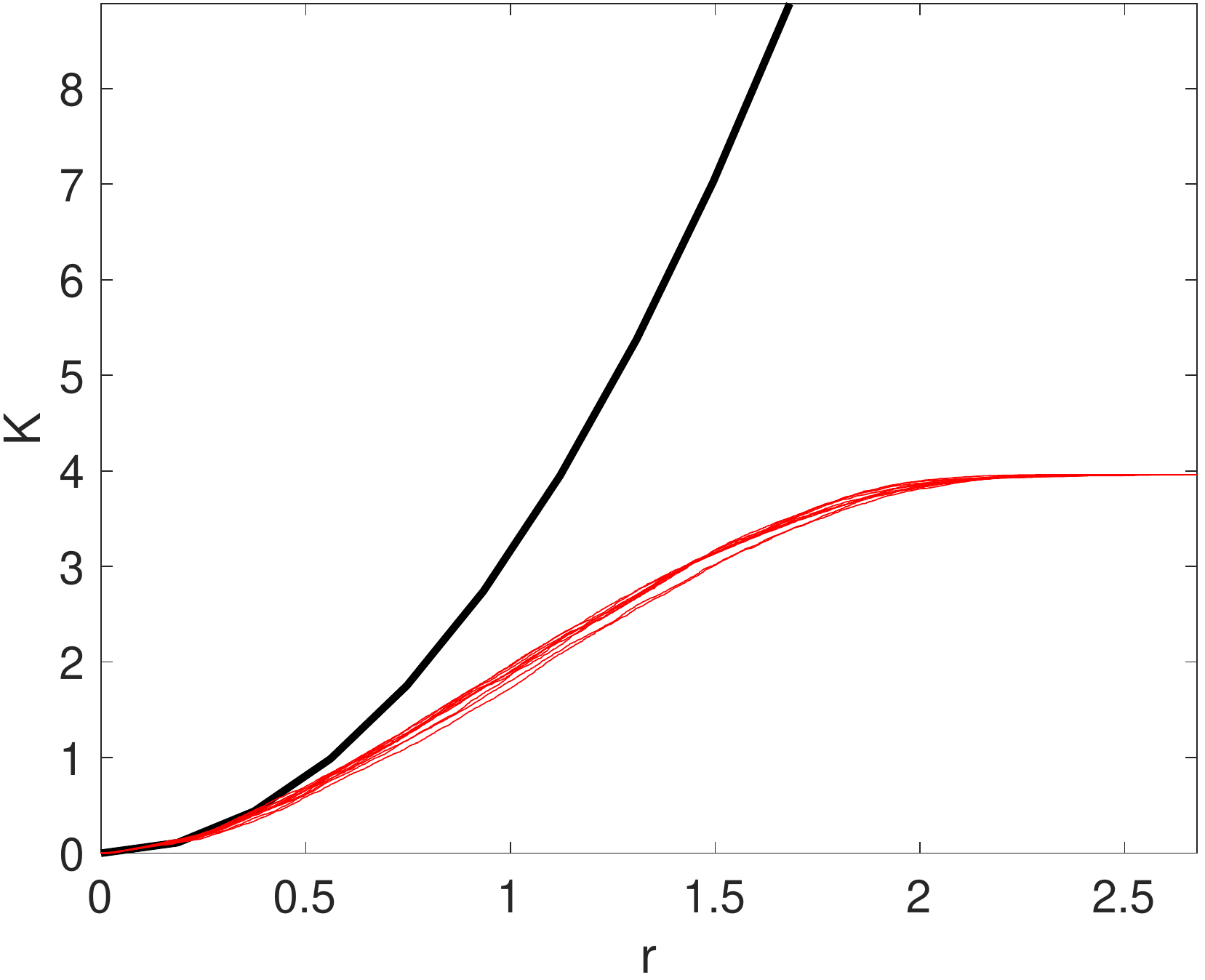}}\hspace{1mm}
  \subfloat[2 narrow clusters]{\label{fig:randomPointsK2Narrow}\includegraphics[width=0.23\linewidth]{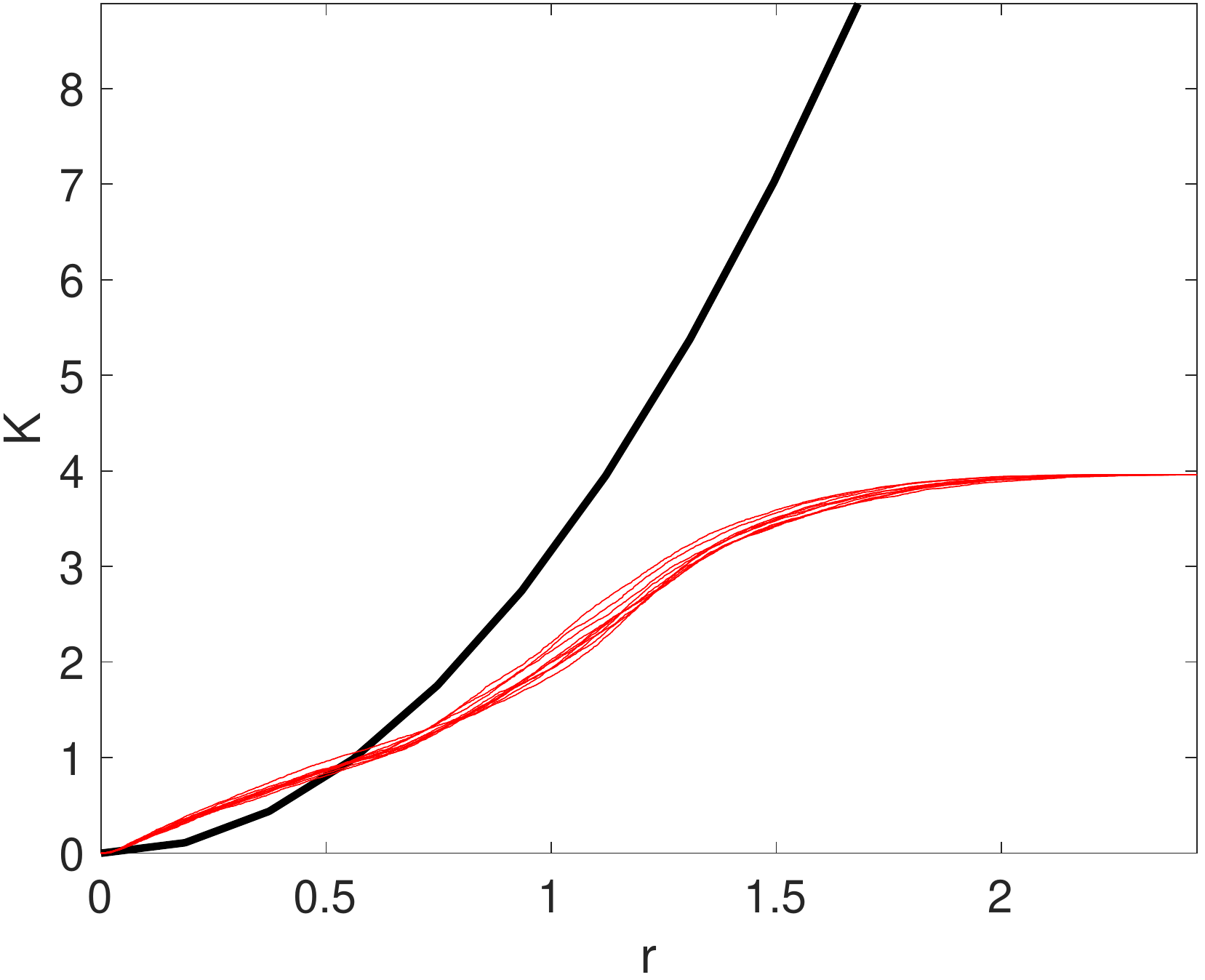}}\hspace{1mm}
  \subfloat[1 narrow cluster]{\label{fig:randomPointsK1Narrow}\includegraphics[width=0.23\linewidth]{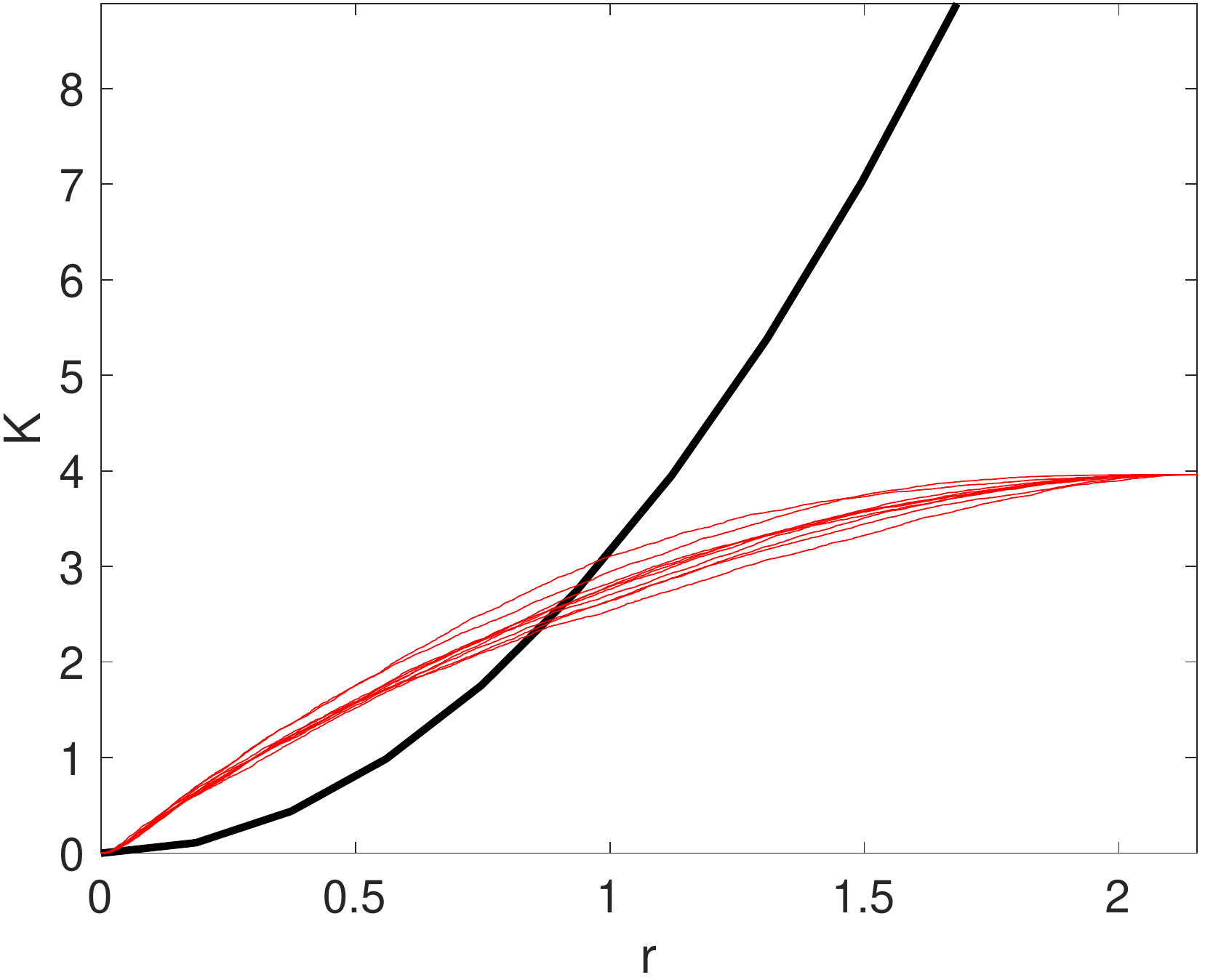}}
  \caption{Repeated estimates of $K_f$ for randomly generated points on randomly generated curves. Black shows the theoretical curve for a Poisson process, and each red curve corresponds to an experiment.}
  \label{fig:randomPointsK}
\end{figure}

To generate random curves in $\Re^2$, we draw a second-degree polynomial $f: \Re^2\rightarrow\Re$ with random coefficients $a_i$, and we choose a distribution $p(x,y)$ of initial points. Then we draw $n$ initial points and develop a curve along the gradient field of $f$. Thus, we consider curve pieces that extends beyond the observation window. Examples of this is shown in \Cref{fig:generatingRandomCurves}. 
A Cox process is
\rwrev{generated by distributing a random number of points uniformly along each curve}. In \Cref{fig:randomPoints} we show four sets of curves, which will be used as comparison cases throughout this paper.
The curves were generated by the same vector field shown in
\Cref{fig:generatingRandomCurves}, but where the distribution of
initial points varies from being essentially uniform \rwrev{or
  distributed within} 7, 2, or 1 localized region. 100 curves were randomly generated for each set. The blue dots show the sample points along the curves, and we use \Cref{lst:sorting} to estimate $K_f$ for these points. The result of 10 experiments, where 100 points were sampled on the same set of curves is shown in \Cref{fig:randomPointsK}.
All estimates of $K_f$ in these experiments seem to converge to 4 for large values of $r$. This is due to the identical sampling and observation window and that no boundary correction terms are included, thereby implying that there are no points outside the observation window. The estimated $K_f$ for the wide and the narrow cluster (\Cref{fig:randomPointsKWide} and \Cref{fig:randomPointsK1Narrow}) shows the expected behavior: For the wide initial set of curves, the $K_f$ initially follows $\pi r^2$, while for the narrow set of curves, the estimated $K_f$ is above $\pi r^2$. The estimates of $K_f$ for 7 \rwrev{regions} is disappointingly similar to the those of the wide cluster, indicating that \rwrev{$K_f$ is not well suited to distinguish the clustering behavior of these} 2 sets of curves. For clusters, there is some structure in the estimated $K_f$. Nevertheless, by these experiments we are motivated to investigate other functions that are similar to $K_f$, but with better descriptive power.

\section{\rwrev{A} $K$-function by Morphology: $K_m$}
\label{sec:morphology}
\rwrev{The $K_f$-function essentially measures aggregated curve length within neighborhoods and does not take into account the geometry of the } curves. Below we present one of two algorithmically inspired alternatives: An essential part of Ripley's $K$-function is the distance function, and the statistical \rwrev{power for distinguishing various clustering or repulsion behavior may be increased by} incorporating other distance or pseudo-distance measures in the \rwrev{construction} of a $K$-function. Thus, for this article, we are less concerned with the measurement theoretic basis and more with the \rwrev{ability of the resulting functions} to distinguish groups of curves. Our first alternative to $K_f$ is to replace the point-wise ball of radius $r$ with a dilation of curves with a ball of radius $r$.

Consider a smooth space curve in $c : \Re \rightarrow \Re^3$ and its length, $\len{c} = \int \norm{c'(t)}\,dt$, where $c'$ is the tangent vector of $c$ and $\norm{c'}$ is its length. The intersection of a curve with a region of interest $W$ may result in several disjoint curves $c_i$, and we introduce the intersection length as,
\begin{equation}
  \label{eq:intersectionLength}
  \intlen{W}{c} = \sum_i\int_{\Gamma_i} \norm{c'(t)}\,dt,
\end{equation}
where we sum over all intervals $\Gamma_i\subset\Re$ for which the curve intersects $W$. We define a curve's $r$-neighborhood by dilation as the set
\begin{equation}
  \label{eq:neighbourhood}
  \mathcal{N}_{r,j} = \mathcal{N}_r(c_j) = c_j \oplus b(0,r),
\end{equation}
for a ball of center 0 and radius $r$. Thus, $\mathcal{N}_{r,j}$ is the set of all points closer than $r$ to the curve. For a set of curves $c_j$, $j=1\ldots n$, we generalize Ripley's $K$-function as,
\begin{equation}
  \label{eq:generalizedRipleysK}
  K_m(r) = \frac{1}{n\rwrev{\rho}}\sum_{j}\frac{1}{\intlen{W}{c_j}}\sum_{i\neq j} \intlen{W\cap\mathcal{N}_{r,j}}{c_i}
\end{equation}
where \rwrev{$\rho$} is the mean curve length per unit area.

The geometry is illustrated in \Cref{fig:geometricComparison:Dilation}. Similarly to \Cref{fig:geometricComparison:Points}, we consider the curve shown as a thick line. The stippled lines parallel to this denotes a dilation with a ball with radius $r$, and the length of the other lines \rwrev{intersection with} the region between the two stippled parallel lines are the curve's contribution to $K_m$. Further, for straight parts of the thick line, the contribution will be close to counting how many neighboring lines there are within a distance $r$ in the normal direction. However, when the thick line has curvature, then neighboring lines are 'counted' in a wedge shape, as illustrated by the two stippled straight lines perpendicular to the thick line.

\subsection{Some Experiments with Random Curves and $K_m$}
\label{sec:ExperimentsMorphology}
\begin{figure}%
  \centering
  \subfloat[Wide]{\label{fig:randomCurvesMorphologyWide}\includegraphics[width=0.23\linewidth]{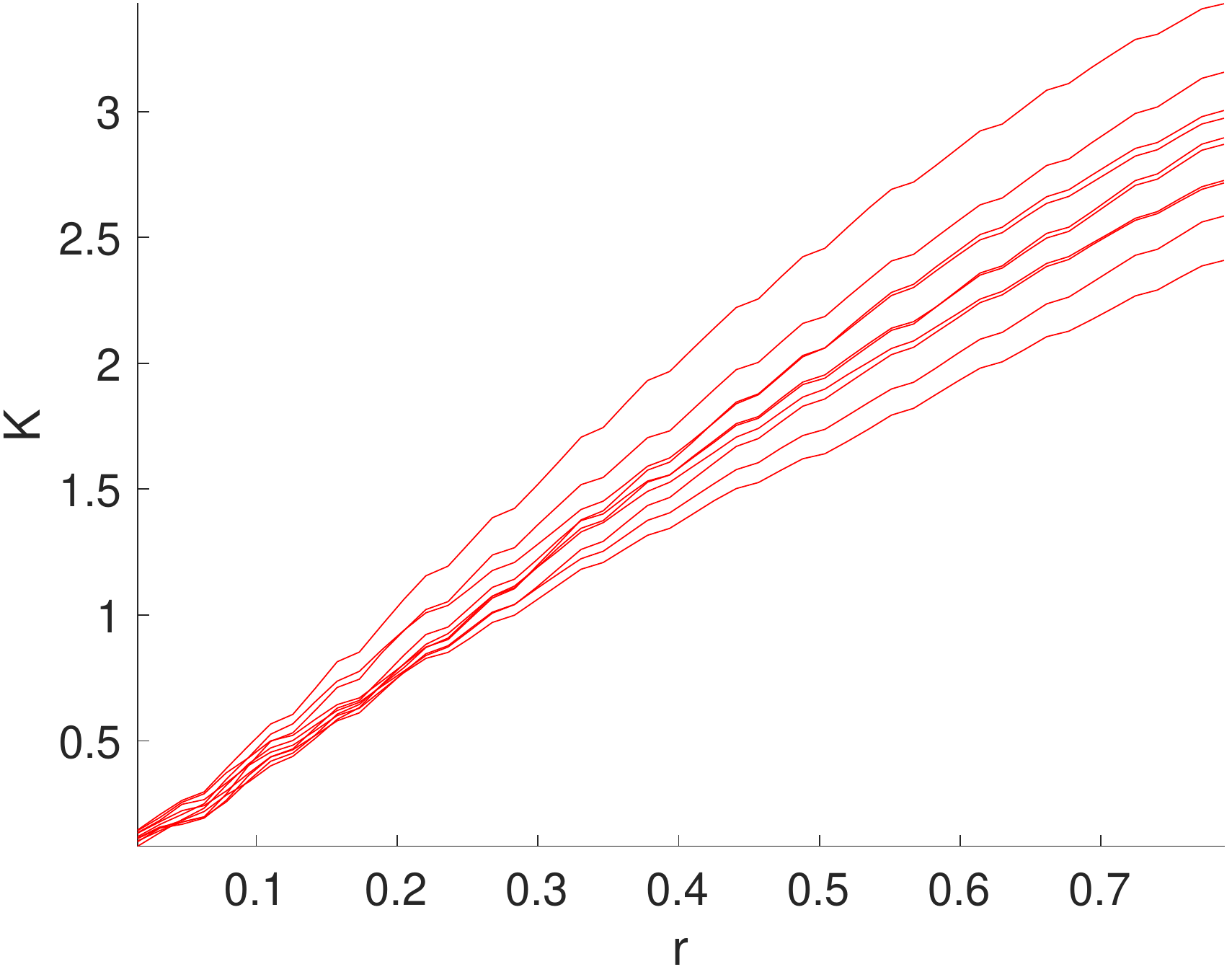}}\hspace{1mm}
  \subfloat[7 narrow clusters]{\label{fig:randomCurvesMorphology7Narrow}\includegraphics[width=0.23\linewidth]{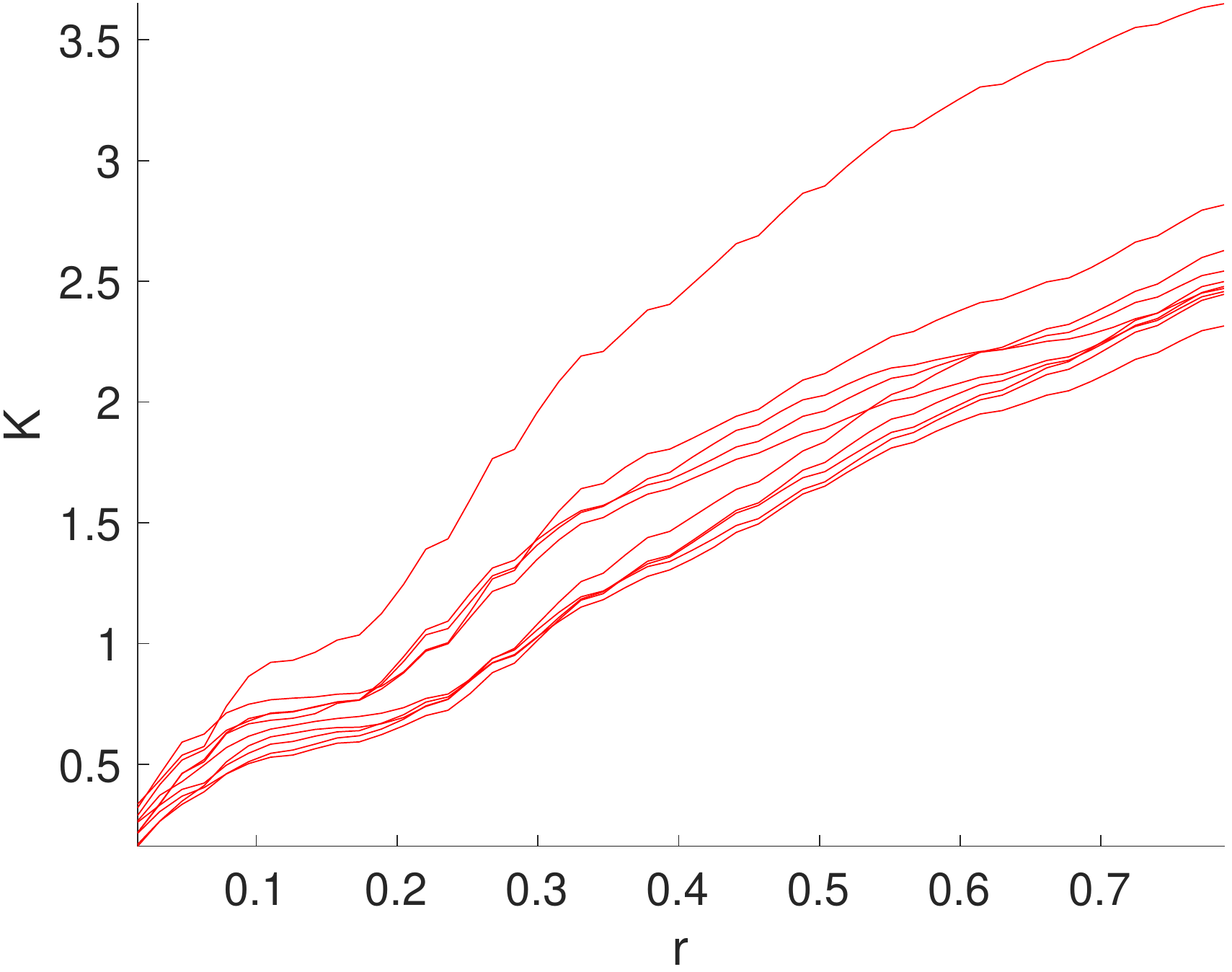}}\hspace{1mm}
  \subfloat[2 narrow clusters]{\label{fig:randomCurvesMorphology2Narrow}\includegraphics[width=0.23\linewidth]{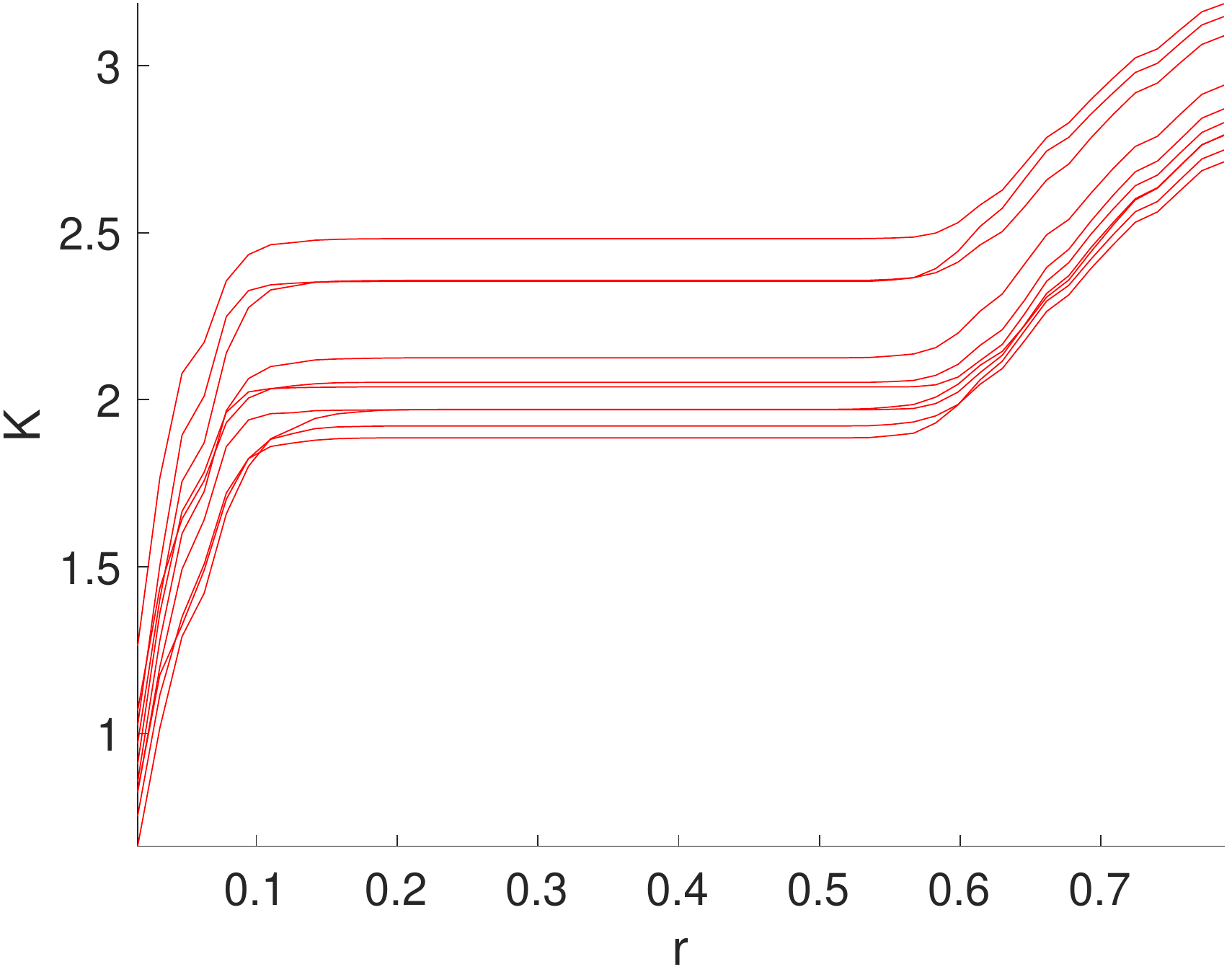}}\hspace{1mm}
  \subfloat[1 narrow cluster]{\label{fig:randomCurvesMorphology1Narrow}\includegraphics[width=0.23\linewidth]{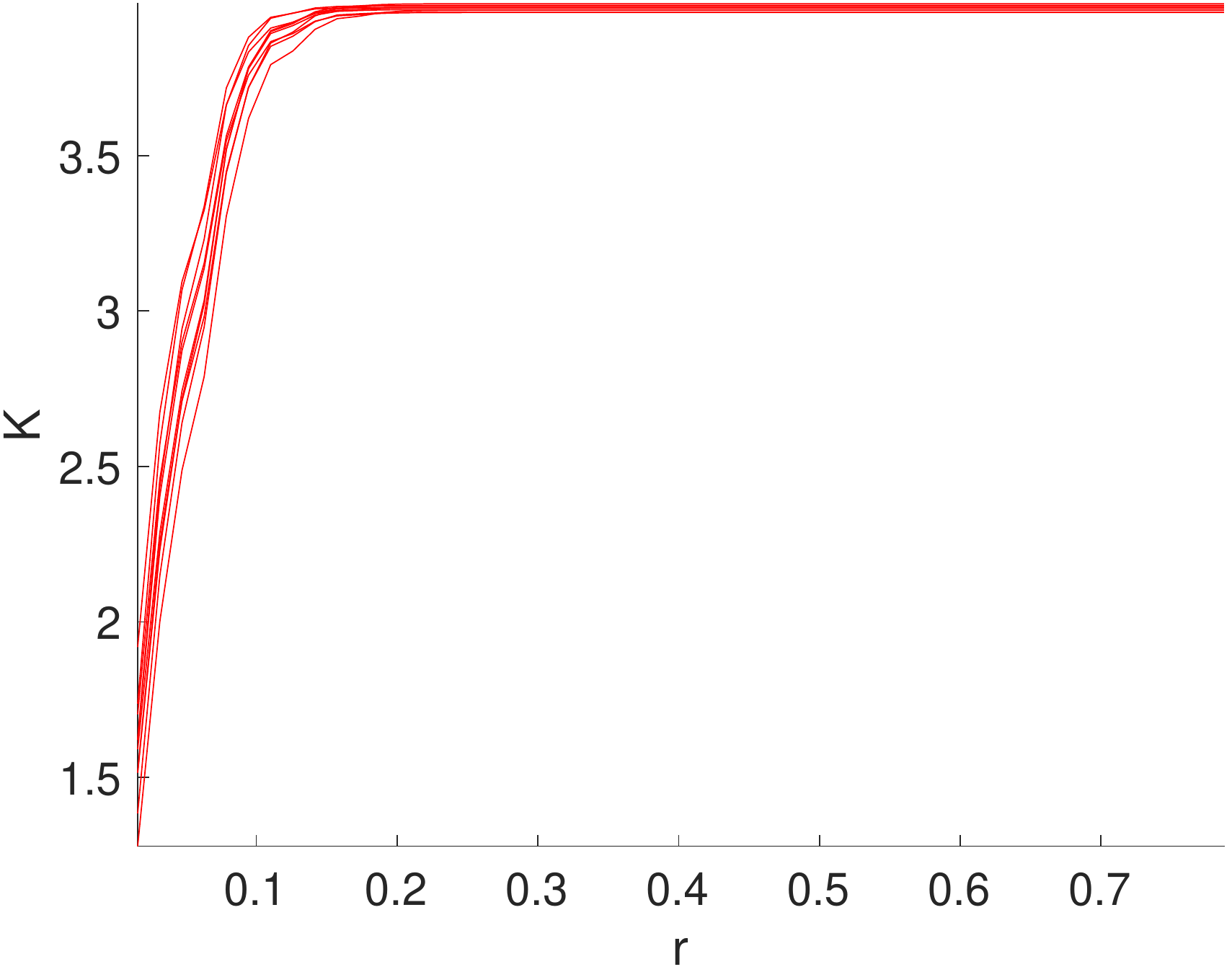}}
  \caption{$K_m$ on randomly generated curves as demonstrated by \Cref{fig:generatingRandomCurves}.}
  \label{fig:randomCurvesMorphology}
\end{figure}

We have performed experiments on the same curves described in \Cref{sec:ExperimentsChiu}. We have performed 10 trials with 30 random curves, and the result is shown in \Cref{fig:randomCurvesMorphology}.
Note that in these experiments, we compare full curves with each other instead of points on curves with each other. The spread of the estimates of the $K_m$-by-dilation functions is greater than in Figure~\ref{fig:randomPointsK}. This is to be expected since \rwrev{the estimates are based} on fewer samples, 30 curves as compared to 100 points. In terms of the shape of $K_m$, we see that \Cref{fig:randomCurvesMorphologyWide} is close to linear for all radii. The remaining experiments show a close to linear growth for $r\in [0,0.1]$, which corresponds well with the width of each individual cluster of curves. Further, \Cref{fig:randomCurvesMorphology7Narrow} shows a plateau for $r \in [0.1,0.2]$, which corresponds well to the distances between its clusters. Similarly, \Cref{fig:randomCurvesMorphology2Narrow} has a plateau for $r\in [0.1,0.6]$, which again is in close relation with the distance between the two bundles. Finally, \Cref{fig:randomCurvesMorphology1Narrow} plateaus for $r>0.1$, which is to be expected, since no other clusters are present in the data. Thus, for these cases, $K_m$ much better separates the different clustering behavior of our curves.

\section{\rwrev{A} $K$-function for Current Metrics: $K_c$}
\label{sec:currents}
In this section, we will define the $K_c$ by currents, where we replace the Euclidean distance in \Cref{lst:sorting} with that of currents for curves.

\subsection{Currents}
The theory of currents, see e.g. \cite{glaunes_transport_2005,durrleman_2010}, embed geometric objects, here specifically curves, into vector spaces that appear as dual spaces to a set of test vector fields. This embedding allows structure from the vector space of test fields to be carried to the space of curves. Specifically, the currents representation inherits the metric structure of the vector space. We here briefly outline the currents construction referring to the above references for details.

Let $l:I\to\Omega$ denote a $C^1$ line in the domain $\Omega\subseteq\mathbb R^d$, and let $v$ be a square integrable vector field on the domain. The assignment $l(v):=\int_l v(x)^T\dot{l}(x)d\lambda(x)$ produces a real number: It is the integral over the line of the inner product between the vector field and the derivative $\dot{l}=\frac{d}{dt}l(t)$ of the curve evaluated at the point $x\in \Omega$. The integral is with respect to the Lebesgue measure on the line. The use of the inner product makes the assignment linear in $v$. If we settle on a suitable space $V$ to which $v$ can belong, then the assignment allows $l$ to be seen as a member of the dual space $V^*$. The currents construction is this embedding of $l$ into $V^*$. Note that though any $C^1$ curve can be embedded in $V^*$ in this way, we cannot expect to find a curve representing e.g. an arbitrary linear combination of elements of $V^*$.

If $V$ is a normed space, we can inherit from it a norm on $V^*$ by setting $\|l\|_{V^*}=\sup_{v\in V, \|v\|=1}|l(v)|$. This gives a metric on the space of curves by setting $d(l_1,l_2)=\|l_1-l_2\|_{V^*}$. To be more explicit, the currents construction assumes $V$ is a reproducing kernel Hilbert space with reproducing kernel $G:\Omega\times\Omega\to\mathbb R^{d\times d}$. Then, for $p\in\Omega$ and $\alpha\in\mathbb R^d$, the map $G(\cdot,p)\alpha$ is a vector field on $\Omega$, and elements of $V$ appear as infinite linear combinations of such elements. The reproducing property implies that $V$ has an inner product $\langle\cdot,\cdot\rangle_V$ and that
\begin{equation}
  \big\langle G(p_1,\cdot)\alpha_1, G(p_2,\cdot)\alpha_2, \big\rangle_V = \alpha_1^T G(p_1,p_2) \alpha_2 \, .
\end{equation}
The inner product defines a norm $\|\cdot\|_V$ and metric $d_V$ on $V$. 

For currents, the interest is rather $V^*$ than $V$. A special set of elements in $V^*$, the Dirac delta currents $\delta_p^\alpha$, take a role in $V^*$ similar to the vector fields $G(\cdot,p)\alpha$ in $V$. Particularly, we can regard $l$ as an infinite sum of such currents with the point $l(t)$ represented by $\delta_{l(t)}^{\dot{l}(t)}$. $V^*$ inherits the inner product from $V$ and $\langle\delta_{p_1}^{\alpha_1},\delta_{p_2}^{\alpha_2}\rangle_{V^*}=\langle G(p_1,\cdot)\alpha_1, G(p_2,\cdot)\alpha_2, \big\rangle_V$. If we discretize $l$ into a finite set of Dirac delta currents $l\approx \sum_{i=1}^n\delta_{l(t_i)}^{\dot{l}(t_i)}$, the current norm of $l$ appears as $\|l\|^2\approx\|\sum_{i=1}^n\delta_{l(t_i)}^{\dot{l}(t_i)}\|^2=\sum_{i,j=1}^n\dot{l}(t_i)^TG(l(t_i),l(t_j))\dot{l}(t_j)$. This norm is consistent when we pass to the limit using an infinite number of points to represent $l$. Note that the points here are oriented because each Dirac delta $\delta_{l(t_i)}^{\dot{l}(t_i)}$ carries the derivative of curve in the vector ${\dot{l}(t_i)}$. The currents can be seen as dual spaces to the set of differential $m$-forms for any $m$. The construction, therefore, applies as well to unoriented points ($m=0$) and surfaces ($m=2$).

The currents norm necessitate a choice of reproducing kernel Hilbert space structure on $V$. An often used choice is the Gaussian kernel $G(x,y)=\alpha e^{-\frac{\|x-y\|^2}{2\sigma^2}}\mathrm{Id}_d$ where $\mathrm{Id}_d$ is the identity matrix on $\mathbb R^d$ and $\alpha,\sigma\in\mathbb R$ the amplitude and variance of the kernel, respectively.

Distance between two discretized curves is defined by the inner product as,
  \begin{align}
    \label{eq:distanceCurrents}
  &d_c(c_1,c_2) = \norm{\sum_iG(p_{1i},\cdot )\alpha_{1i}-\sum_jG(p_{2j},\cdot )\alpha_{2j}}\\\nonumber
  &\quad= \sqrt{\sum_i\sum_j\alpha_{1i}^T G(p_i,p_j) \alpha_{1j}+\sum_i\sum_j\alpha_{2i}^T G(p_i,p_j) \alpha_{2j}-2 \sum_i\sum_j\alpha_{1i}^T G(p_i,p_j) \alpha_{2j}}.
\end{align}
Finally, we define $K_c$ as \Cref{lst:sorting} using $d(c_1,c_2) = \min\left(d_c(c_1,c_2),d_c(c_1,-c_2)\right)$, where $-c_2$ is the curve with orientation opposite of $c_2$, see also the varifold representation \cite{charon.trouve13}.

The geometry is illustrated in \Cref{fig:geometricComparison:Currents}. Similarly to \Cref{fig:geometricComparison:Points}, we consider a step in the process of estimating the $K_c$: a Gaussian, depicted as a ball, has been placed around the point on a curve (thick line) with its tangent vector at this point shown as a blue arrow. We then consider other points and their tangent vectors, and the dot product of the tangent at the origin and at the other points are in turn calculated and weighted by their distance according to $G$. As a consequence, the current distant measure emphasizes points on curves in a soft neighborhood that are close to parallel.

\subsection{Some Experiments with Random Curves and $K_c$}
\begin{figure}%
  \centering
  \subfloat[Wide]{\includegraphics[width=0.23\linewidth]{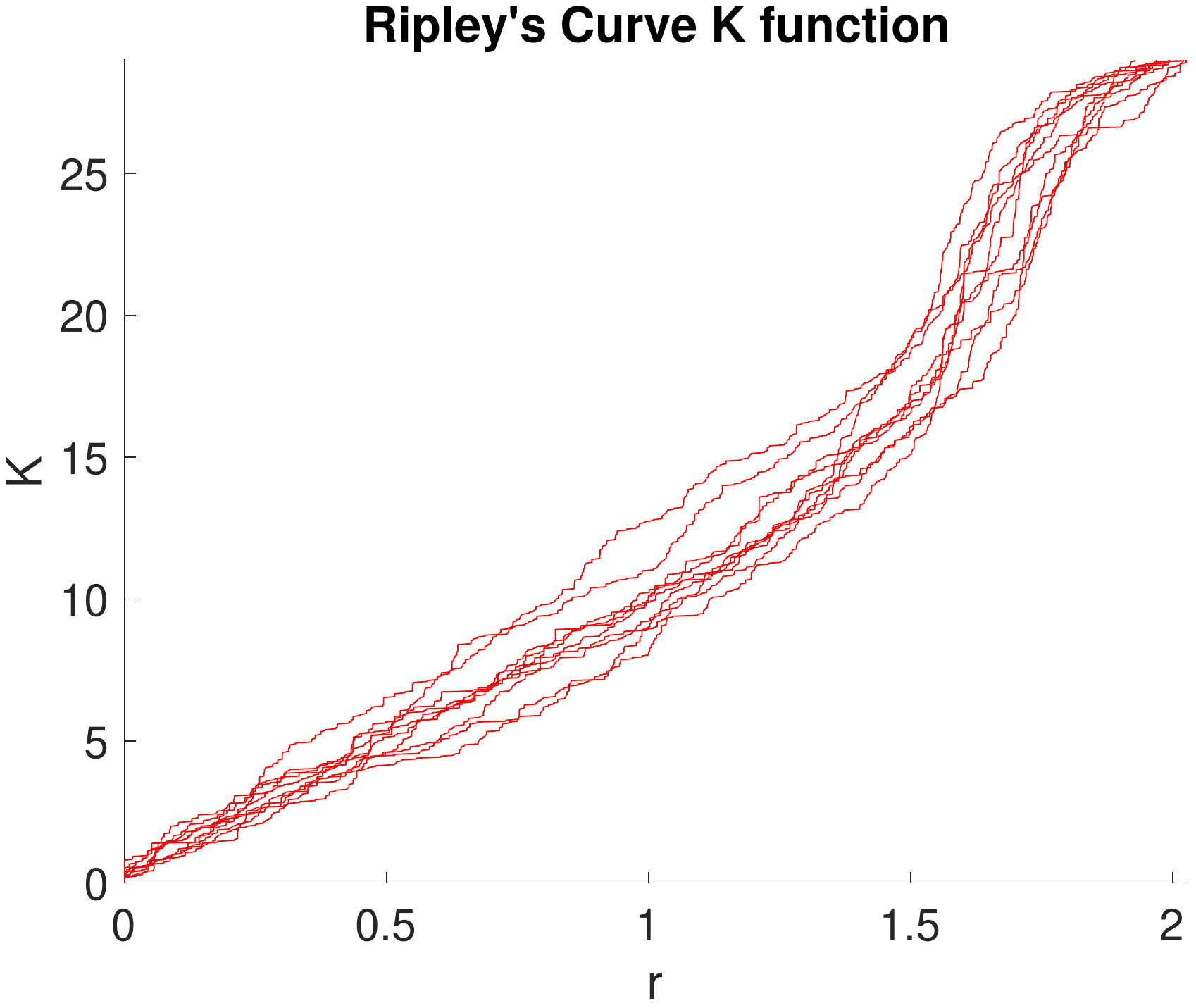}}\hspace{1mm}
  \subfloat[7 narrow clusters]{\includegraphics[width=0.23\linewidth]{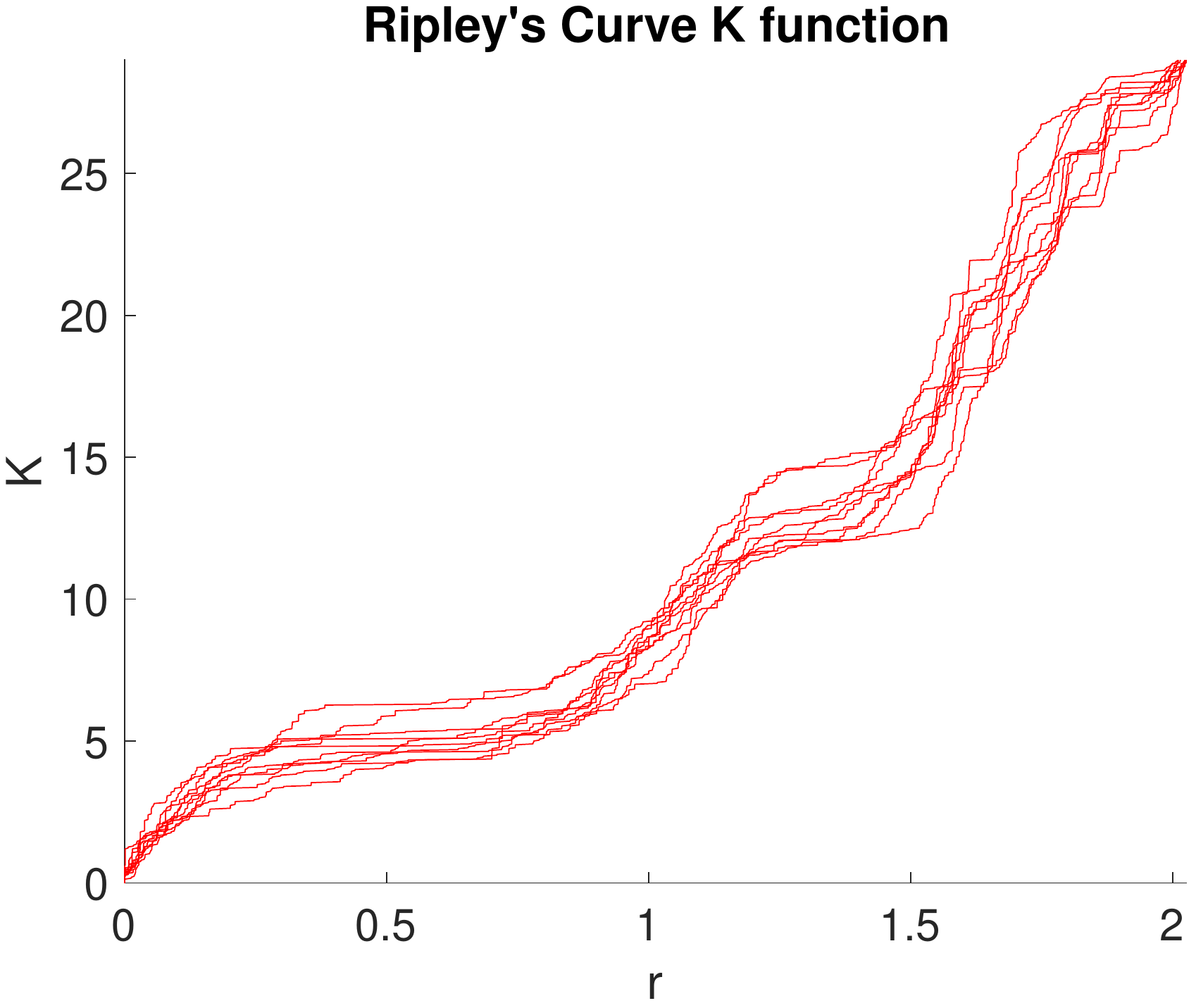}}\hspace{1mm}
  \subfloat[2 narrow clusters]{\includegraphics[width=0.23\linewidth]{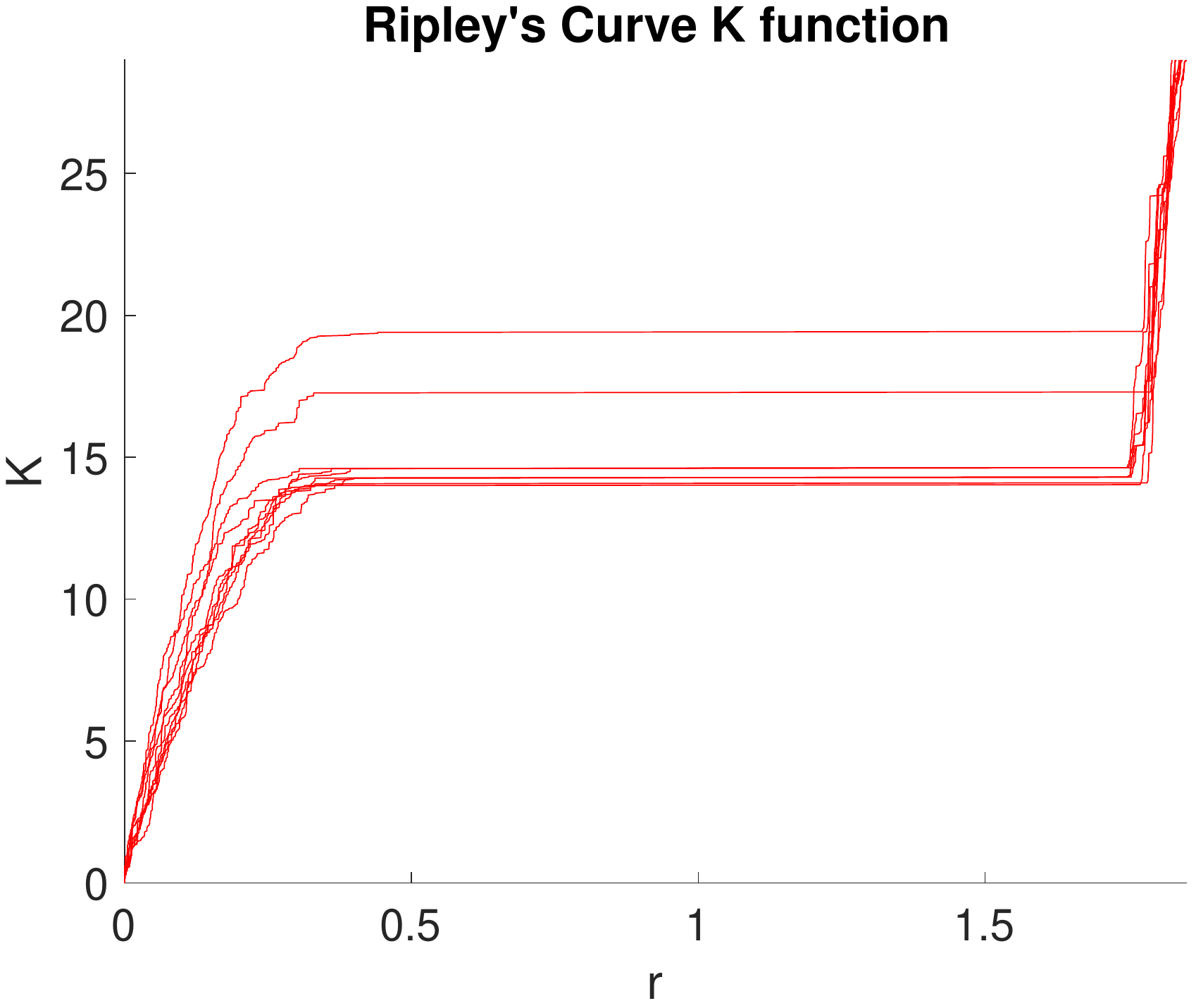}}\hspace{1mm}
  \subfloat[1 narrow cluster]{\includegraphics[width=0.23\linewidth]{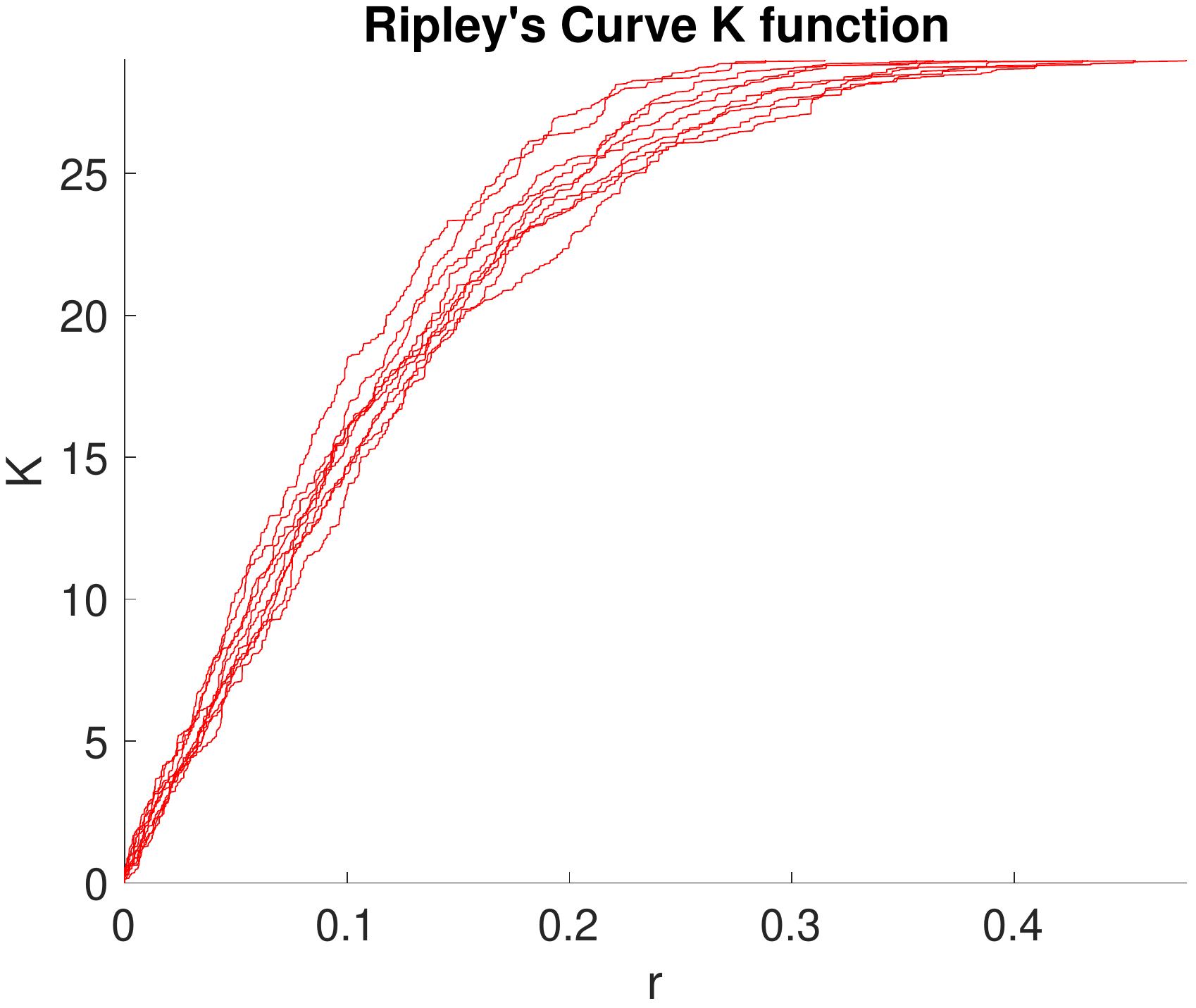}}
  \caption{$K_c$ using $\sigma=0.5$ and on curves as demonstrated by \Cref{fig:generatingRandomCurves}.}
  \label{fig:randomCurvesCurrents05}
\end{figure}

We have performed experiments on the same curves as described in \Cref{sec:ExperimentsChiu}. As in \Cref{sec:ExperimentsMorphology} we performed 10 trials with 30 random curves from the same random set of curves described in \Cref{sec:ExperimentsChiu}. We have used the Gaussian kernel with $\sigma = 0.5$ to correspond with the expected size of statistical structures observed in the data. The result is shown in \Cref{fig:randomCurvesCurrents05}.
%
%
As the experiments with $K_m$, all curves show a linear increase starting at $r=0$, and superficially, the experiment with 7, 2, and 1 narrow cluster show a stepping like behavior similarly to $K_m$. However, interpretation of these curves is slightly more complicated, since they describe a mixture of pointwise distance and the dot product of their tangents. Curves that are parallel will have small values of \eqref{eq:distanceCurrents}, while curves that are perpendicular will have large distances.

\section{Discussion and conclusion}
\label{sec:conclusion}
In this article, we have considered Ripley's \rwrev{$K$-function for sets of points and a $K$-function, $K_f$, for curves}, and we observed that the two definitions are very similar \rwrev{for a certain type of Cox} processes. We also observed that the definition of $K_f$, given in \cite{chiu:stoyan:kendall:mecke:13}, is not well suited to distinguish simple line patterns, as demonstrated on a small number of cases. This has motivated us to consider two alternative $K$-like functions: $K_m$, which uses mathematical morphology to define pseudo-distance between curves, and $K_c$, which defines distances by currents. Both give rise to $K$-functions that are more descriptive in terms of the simple examples presented. However, they differ in how they interpret the geometry of curves and their relations. Oversimplified examples may shed some light on their differences: Consider two infinite lines in an infinite two-dimensional space crossing at a right angle. The contribution for a point right at the center will for the three $K$-functions be: $K_f(r) = \mathcal{O}(r)$, $K_m(r) = \frac{r}{L}$, $K_c(r) = 0$. In contrast, for two infinite parallel lines at distance $d$, for a point on one of the lines, the three $K$-functions will be $K_f(r) = 0 \text{ if } r < d \text{ else } \mathcal{O}(\sqrt{r^2+d^2})$, $K_m(r) = 0 \text{ if } r < d \text{ else } 1$, or $K_c(r) \simeq 0 \text{ if } r < \sigma\sqrt{2\pi}\exp{\frac{-d^2}{2\sigma^2}} \text{ else } 1$. Thus, for orthogonal lines, $K_f$ and $K_m$ are similar, while for parallel lines, $K_m$ and $K_c$ \rwrev{are similar}. All three $K$-functions are highly general, easily defined in higher dimensions and for manifold pieces of higher dimensions. However, it is expected that $K_f$ will be the fastest to compute since it only depends on sampled points on manifold pieces. $K_c$ is also fast to compute for rotational symmetric kernels, which likewise can be approximated by sample points and their tangent space. However, $K_c$ has a free parameter, which require further investigation in order to properly set.  The pseudo-distance measure used for $K_m$ has geometrical appeal to these authors, and may have useful applications.


Our future work will focus on studying the relation between the (pseudo-) distance function used, the implied $K$-function, and the statistical problem to be solved.

\bibliographystyle{unsrt}
\bibliography{paper}

\begin{thebibliography}{1}

\bibitem{chiu:stoyan:kendall:mecke:13}
S.N. Chiu, D.~Stoyan, W.S. Kendall, and J.~Mecke.
\newblock {\em Stochastic Geometry and Its Applications}.
\newblock Wiley Series in Probability and Statistics. Wiley, 2013.

\bibitem{ripley77}
B.~D. Ripley.
\newblock Modelling spatial patterns.
\newblock {\em Journal of the Royal Statistical Society. Series B
  (Methodological)}, 39(2):172--212, 1977.

\bibitem{baddeley:rubak:turner:15}
A.~Baddeley, E.~Rubak, and R.~Turner.
\newblock {\em Spatial point patterns: methodology and applications with {R}}.
\newblock Chapman and Hall/CRC, 2015.

\bibitem{Durrleman.Pennec.Trouve.Ayache09}
Stanley Durrleman, Xavier Pennec, Alain Trouv\'{e}, and Nicholas Ayache.
\newblock Statistical models of sets of curves and surfaces based on currents.
\newblock {\em Medical Image Analysis}, 13:793–808, 2009.

\bibitem{glaunes_transport_2005}
Joan Glaun\`es.
\newblock {\em Transport Par Diff\'eomorphismes de Points, de Mesures et de
  Courants Pour La Comparaison de Formes et l'anatomie Num\'erique}.
\newblock PhD thesis, Universit\'e Paris 13, Villetaneuse, France, 2005.

\bibitem{durrleman_2010}
Stanley Durrleman.
\newblock {\em Statistical models of currents for measuring the variability of
  anatomical curves, surfaces and their evolution}.
\newblock PhD thesis, de l’Université Nice - Sophia Antipolis, France, 2010.

\bibitem{charon.trouve13}
N.~Charon and A.~Trouvé.
\newblock The varifold representation of nonoriented shapes for diffeomorphic
  registration.
\newblock {\em SIAM Journal on Imaging Sciences}, 6(4):2547--2580, 2013.

\end{thebibliography}
\end{document}